\documentclass{aa}

\usepackage{comment}
\usepackage{ulem}
\usepackage{color}
\usepackage{graphicx}
\usepackage{txfonts}
\usepackage{lipsum}
\usepackage[]{hyperref}
\hypersetup{colorlinks=true,linkcolor=black,citecolor=blue}

\def\vec#1{\mbox{\boldmath $#1$}}

\usepackage[dvipsnames]{xcolor}

\begin{document}

    \title{Assessing the capability of a model-based stellar XUV estimation}
    
    \author{Munehito Shoda
    \inst{1},
    Kosuke Namekata
    \inst{2}
    \and
    Shinsuke Takasao
    \inst{3}
    }
    
    \institute{
    Department of Earth and Planetary Science, School of Science, The University of Tokyo, 7-3-1 Hongo, Bunkyo-ku, Tokyo, 113-0033, Japan
    \email{shoda.m.astroph@gmail.com}
    \and
    National Astronomical Observatory of Japan, 
    National Institutes of Natural Sciences, 2-21-1 Osawa, Mitaka, Tokyo, 181-8588, Japan
    \and
    Department of Earth and Space Science, Graduate School of Science, Osaka University, Toyonaka, Osaka 560-0043, Japan
    }
    
    \date{Received month dd, yyyy; accepted month dd, yyyy}

    \abstract
    {
    Stellar XUV (X-ray and extreme ultraviolet) emission drives the heating and chemical reactions in planetary atmospheres and protoplanetary disks, and therefore, a proper estimation of a stellar XUV spectrum is required for their studies. One proposed solution is to estimate stellar atmospheric heating using numerical models, although the validation was restricted to the Sun over a limited parameter range. In this study, we extend the validation of the model by testing it with the Sun and three young, nearby solar-type stars with available XUV observational data ($\kappa^1$ Ceti, $\pi^1$ UMa and EK Dra). We first test the model with the solar observations, examining its accuracy in activity minimum and maximum phases, its dependence on loop length, the effect of loop length superposition, and its sensitivity to elemental abundance. We confirm that the model spectrum is mostly accurate both in activity minimum and maximum, although the high-energy X-rays ($\lambda < 1 {\rm \ nm}$) are underestimated in the activity maximum. Applying the model to young solar-type stars, we find that it can reproduce the observed XUV spectra within a factor of 3 in the range of 1-30 nm for stars with magnetic flux up to 100 times that of the Sun ($\kappa^1$ Ceti and $\pi^1$ UMa). For a star with 300 times the solar magnetic flux (EK Dra), although the raw numerical data show a systematically lower spectrum than observed, the spectra are in good agreement once corrected for the effect of insufficient resolution in the transition region. For all young solar-type stars, high-energy X-rays ($\lambda < 1 {\rm \ nm}$) are significantly underestimated, with the deviation increasing with stellar magnetic activity. Furthermore, our model-based estimation shows performance that is comparable to or surpassed that of previous empirical approaches. We also demonstrate that the widely used fifth-order Chebyshev polynomial fitting can accurately reproduce the actual DEM and XUV spectrum. Our findings indicate that the stellar XUV spectrum can be reasonably estimated through a numerical model, given that the essential input parameters (surface magnetic flux and elemental abundance) are known.
    }
    

    \titlerunning{XUV model of solar-type stars}
    \authorrunning{M. Shoda, K. Namekata and S. Takasao}
    
    \maketitle
    
\section{Introduction}

Low-mass stars possess intrinsic magnetic fields \citep{Landstreet_1992_AARv, Berdyugina_2005_LRSP, Reiners_2012_LRSP, Brun_2017_LRSP, Isik_2023_SSRv}, serving as the energy source for various magnetic activities.A phenomenon induced by magnetic field includes coronal heating \citep{Klimchuk_2006_SolPhys, Aschwanden_2007_ApJ, Parnell_2012_RSPTA, De_Moortel_2015_RSPTA} and the emission of XUV radiation from the heated corona \citep{Dere_1982_SoPh, Moses_1997_SoPh, Del_Zanna_2011_AA}. Stellar XUV photons can heat planetary atmospheres and trigger chemical reactions \citep{Lammer_2003_ApJ, Kulikov_2007_SSRev, Lecavelier_des_Etangs_2007_AA, Stone_2009_ApJ, Terada_2009_AsBio, Garcia-Sage_2017_ApJ, Mitani_2022_MNRAS, Yoshida_2022_ApJ, Nishioka_2023_JGR}, making it an important factor in the evolution of planetary atmospheres. They also play a crucial role in the evolution of protoplanetary disks during the pre-main sequence stage \citep{Gorti_2009_ApJ_photoevaporation_rate, Gorti_2009_ApJ_time_evolution_of_PPD, Nakatani_2018_ApJ, Ercolano_2021_MNRAS, Kunitomo_2021_ApJ, Nakatani_2023_ApJ, Nakatani_2024_arXiv}. Thus, understanding a stellar XUV emission spectrum is essential for quantitatively describing the evolution of surrounding planets and protoplanetary disks. 

One of the most significant challenges regarding stellar XUV radiation is its observational difficulty. Extreme ultraviolet photons, which constitute most of the XUV emission, are strongly absorbed by the interstellar medium \citep{Cruddace_1974_ApJ, France_2019_SPIE}, making it impossible to observe the full XUV spectrum of stars other than the Sun. While some stars have observable XUV spectra at shorter wavelengths \citep[$\lambda < 40 {\rm \ nm}$,][]{Ribas_2005_ApJ, Ribas_2016_AA, Ribas_2017_AA}, these are limited in number. Consequently, we need to estimate a stellar XUV spectrum from other observable quantities.

Several methods exist for estimating the XUV spectrum of stars. The simplest way is to scale the entire solar XUV spectrum according to the stellar-to-solar X-ray luminosity ratio ($L_{\rm X}/L_{{\rm X}, \odot}$), a technique employed in the literature \citep{Penz_2008_PSS, Tian_2009_ApJ}. However, this method eventually fixes the ratio of X-ray luminosity ($L_{\rm X}$) to EUV luminosity ($L_{\rm EUV}$), which is not always accurate. In reality, $L_{\rm X}/L_{\rm EUV}$ tends to decrease as $L_{\rm X}$ increases \citep{Ribas_2005_ApJ, Sanz-Forcada_2011_AA, Chadney_2015_Icar, Johnstone_2021_AA_XUV}, indicating that the direct scaling of the solar spectrum may not be accurate. Alternatively, XUV flux estimations using correlations with other physical parameters have been proposed. These methods involve estimations based on stellar age \citep{Claire_2012_ApJ}, Lyman-alpha flux \citep{Linsky_2014_ApJ}, or surface magnetic flux \citep{Namekata_2023_ApJ}. In particular, the Lyman-alpha method is widely used in the literature and is incorporated in the comprehensive survey of the spectral energy distribution \citep[MUSCLES Treasury Survey,][]{France_2016_ApJ, Youngblood_2016_ApJ, Loyd_2016_ApJ}.

One of the most sophisticated models for estimating the XUV spectrum is based on the differential emission measure (DEM) \citep{Sanz-Forcada_2011_AA, Chadney_2015_Icar, Louden_2017_MNRAS, France_2020_AJ, Diamond-Lowe_2021_AJ, Duvvuri_2021_ApJ, Wilson_2021_ApJ, Fuhrmeister_2022_AA, Duvvuri_2023_AJ}. DEM represents the emission of plasma along the line of sight as a function of temperature and is defined as follows \citep{Pye_1978_AA, Brueckner_1983_ApJ, Del_Zanna_2001_AA, Landi_2008_ApJ, ODwyer_2011_AA}
\begin{align}
    \xi (T) = n_e^2 \frac{dl_{\rm los}}{dT}, \label{eq:DEM_definition}
\end{align}
where $n_e$ is the number density of electron and $l_{\rm los}$ is the length along the line of sight. The intensity of an optically thin emission line at wavelength $\lambda$, represented by $I_\lambda$, can be expressed using DEM as follows. 
\begin{align}
    I_\lambda = \int dT \ G_\lambda (n_e, T) \xi (T), \label{eq:intensity_dem}
\end{align}
where $G_\lambda (n_e, T)$ is the contribution function at wavelength $\lambda$, which can be computed using atomic data for a known plasma abundance. DEM can be determined by solving the inversion problem of Eq.~\eqref{eq:intensity_dem} for observable line intensities, by imposing some empirical constraints \citep{Gudel_2004_AARv, Schmitt_2004_AA}. Subsequently, applying Eq.~\eqref{eq:intensity_dem} to all wavelengths using the obtained DEM allows for the reconstruction of the spectrum across any wavelength range. 

The DEM-based estimation of the XUV spectrum was first attempted in the generation of the solar reference spectrum \citep{Warren_1998_JGR}. Measuring DEM for stars, with little to no EUV data, is challenging compared to the Sun. However, DEM is partially measurable through X-ray and far-ultraviolet observations. X-ray observations can constrain the high-temperature side of DEM \citep[$T \gtrsim 10^{6.1} {\rm \ K}$,][]{Peres_2000_ApJ, Sanz-Forcada_2003_AA, Sanz-Forcada_2011_AA, Orlando_2017_AA, Fuhrmeister_2022_AA}, while FUV observations can constrain the lower temperature side \citep[$T \lesssim 10^{5.3} {\rm \ K}$,][]{Kjeldseth_Moe_1977_ApJ, Dupree_1993_ApJ, Osten_2006_ApJ, Louden_2017_MNRAS, Diamond-Lowe_2021_AJ, Pineda_2021_ApJ, Fuhrmeister_2022_AA}, allowing for the reconstruction of DEM in the whole temperature range responsible for EUV emissions. Recently, this approach has become common and has been validated by the solar observation \citep{Duvvuri_2021_ApJ}.

While reconstructing XUV spectra using DEM observation is effective, it remains resource-intensive because it requires both X-ray and FUV observations. An alternative solution is to obtain DEM using numerical simulations. Recent advances in numerical models of solar corona heating have significantly enhanced our understanding of the necessary physical processes and required resolution to replicate the solar atmosphere realistically \citep{Gudiksen_2005_ApJ, Rappazzo_2008_ApJ, Bradshaw_2013_ApJ, Hansteen_2015_ApJ, Dahlburg_2016_ApJ, Rempel_2017_ApJ, Breu_2022_AA, Kuniyoshi_2024_ApJ_synthesis_of_magnetic_tornado, Shi_2024_ApJ}. Based on these insights, we developed a numerical model capable of calculating DEM to replicate the solar XUV spectrum \citep[][hereafter Paper I]{Shoda_2021_AA}. A major advantage of this model is that its input parameters, the surface unsigned magnetic flux and elemental abundance, are directly observable via Zeeman broadening \citep{Gray_1984_ApJ, Saar_1996_proceedings, Johns-Krull_1999_ApJ, Saar_2001_proceedings, Reiners_2009_ApJ, Reiners_2022_AA, Hahlin_2023_AA} and spectroscopy \citep{Stelzer_2004_AA, Maggio_2007_ApJ, Testa_2010_SSRev}, respectively. However, the model was validated only for the Sun in the activity minimum, and its applicability to other stars remains uncertain.

Although full-wavelength XUV spectral observation is currently impractical for stars beyond our solar system, we can observe the XUV spectra at shorter wavelengths for a subset of solar-type stars with considerably large magnetic flux. This study aims to evaluate the potential for estimating the XUV spectrum of solar-type stars by comparing simulated and observed XUV spectra for these targets. To this end, the applicability and uncertainty of the model are assessed with respect to the Sun. Once the uncertainty level is determined, the model spectra of solar-type stars are compared with observational data to evaluate the capability of the model.

\section{Model description \label{sec:model}} 

\subsection{Simulation targets and observations}

This study focuses on solar-type stars, specifically main-sequence stars with mass and radius comparable to the Sun. As we compare models and observations based on the XUV spectrum, the simulation targets must meet two criteria: 1) observed surface magnetic flux and elemental abundance (as model input parameters), and 2) observed XUV spectra (to be compared with the model outputs). Based on these criteria, we selected the Sun, $\kappa^1$ Ceti, $\pi^1$ UMa, and EK Dra for our model calculations (see Table~\ref{table:target_parameters} for the fundamental parameters of these targets).

For the observation of the solar XUV spectrum during the solar activity minimum, we utilize data collected throughout the Whole Heliosphere Interval \citep[WHI, from March 20, 2008 to April 16, 2008,][]{Chamberlin_2009_GRL, Woods_2009_GRL} from March 20, 2008, to April 16, 2008, a period characterized by minimal solar activity. For the solar activity maximum, the XUV spectrum is obtained by combining measurements from TIMED/SEE \citep[][$\lambda \le 6 {\rm \ nm}$]{Woods_2005_JGR_TIMED_SEE} and SDO/EVE \citep[][$\lambda > 6 {\rm \ nm}$]{Woods_2012_SoPh}. We use the 1-month averaged data over April 2014, a period characterized by maximal solar activity. For the other stars, combined observations from ASCA, ROSAT, and EUVE \citep{Ribas_2005_ApJ} are retrieved. The solar surface magnetic flux measurements are taken from \citet{Cranmer_2017_ApJ}, while for the other stars, values are derived from the Zeeman broadening analysis \citep{Kochukhov_2020_AA}. The observed surface magnetic fluxes are listed in Table~\ref{table:target_parameters}.

\renewcommand{\arraystretch}{1.5}
\begin{table*}[t!]
\centering
  {\tabcolsep=1.3em
  \begin{tabular}{cccccccc}
    Name
    & $R_\ast$ [$R_\odot$]
    & $\log g$
    & $T_{\rm eff}$ [K]
    & $M_\ast$ [$M_\odot$]
    & $L_\ast$ [$L_\odot$]
    & $B_\ast f_\ast$ [kG]
    & $\Phi_\ast$ [$10^{25}$ Mx]
    \\ \hline \hline
    Sun (min)
    & 1.00
    & 4.44
    & 5777
    & 1.00
    & 1.00
    & 5.4 $\times 10^{-3}$$^{(1)}$
    & 3.3 $\times 10^{-2}$ \\
    Sun (max)
    & 1.00
    & 4.44
    & 5777
    & 1.00
    & 1.00
    & 9.0 $\times 10^{-3}$$^{(1)}$
    & 5.5 $\times 10^{-2}$ \\
    $\kappa^1$ Ceti
    & 0.97
    & 4.46$^{(2)}$
    & 5749$^{(2)}$
    & 1.00$^{(2)}$
    & 0.93
    & 0.50$^{(4)}$
    & 2.9 \\
    $\pi^1$ UMa
    & 1.00
    & 4.44$^{(2)}$
    & 5873$^{(2)}$
    & 1.00$^{(2)}$
    & 1.06
    & 0.59$^{(4)}$
    & 3.6 \\
    EK Dra
    & 1.07
    & 4.40$^{(3)}$
    & 5770$^{(3)}$
    & 1.04$^{(3)}$
    & 1.14
    & 1.4$^{(4)}$
    & 9.8 \\
    \hline \hline
  \end{tabular}
  }
  \vspace{0.5em}
  \caption{Target objects in this study (the Sun at activity minimum and maximum, $\kappa^1$ Ceti, $\pi^1$ UMa, EK Dra) and their observed parameters. The second column shows the stellar radius $R_\ast$ (in solar radii), the third column shows the logarithm (base 10) of the surface gravity $g = GM_\ast/R_\ast^2$ (in cgs unit), the fourth column shows the effective temperature $T_{\rm eff}$, the fifth column presents the stellar mass $M_\ast$ (in solar mass) calculated by $\log g$ and $R_\ast$, the sixth column displays the luminosity $L_\ast = 4 \pi R_\ast^2 \sigma T_{\rm eff}^4$ (in solar luminosity), the seventh column shows the average surface magnetic field strength $\langle B \rangle_\ast$, and the eighth column presents the unsigned surface magnetic flux $\Phi_\ast = 4 \pi R_\ast^2 B_\ast f_\ast$. References: (1) \citet{Cranmer_2017_ApJ}, (2) \citet{Gonzalez_2010_MNRAS}, (3) \citet{Senavci_2021_MNRAS}, (4) \citet{Kochukhov_2020_AA}.}
  \label{table:target_parameters}
\end{table*}

\subsection{Stellar surface parameters}

In this study, we focus on stars with luminosity ($L_\ast$), mass ($M_\ast$), and radius ($R_\ast$) similar to or identical to the Sun. For simplicity, in our numerical simulations, we set these values to be the same as those of the Sun, i.e., 
\begin{align}
    L_\ast = L_\odot, \ \ \ \ M_\ast = M_\odot, \ \ \ \ R_\ast = R_\odot,
\end{align}
implying that the surface temperature is also identical to the solar value: 
\begin{align}
    T_\ast = T_\odot.
\end{align}
We also assume that the metallicity of the photosphere $Z_\ast$ is the same for all the targets, that is,
\begin{align}
    Z_\ast = Z_\odot.
\end{align}

Stellar surface properties can be approximately specified by $L_\ast$, $M_\ast$, $R_\ast$, and $Z_\ast$ \citep{Cranmer_2011_ApJ}. As these parameters are set to be identical to the solar parameters, we set the stellar surface parameters (mass density, convective correlation length, convective frequency, and Poynting flux) to the solar values:
\begin{align}
    &\rho_\ast = 1.87 \times 10^{-7} {\rm \ g \ cm^{-3}}, \ \ \ \ \lambda_{\perp,\ast} = 1.50 \times 10^{7} {\rm \ cm}, \nonumber \\
    &f_\ast^{\rm conv} = 1.00 \times 10^{-3} {\rm \ s^{-1}}, \ \ \ \ F_{A,\ast} = 2.40 \times 10^9 {\rm \ erg \ cm^{-2} \ s^{-1}},
\end{align}
where $\lambda_{\perp,\ast}$ is the typical diameter of the magnetic bright points embedded within the intergranular lanes \citep{Wiehr_2004_AA, Utz_2009_AA, Xiong_2017_ApJ, Berrios_Saavedra_2022_AA, Kuniyoshi_2023_ApJ} and $1/f_\ast^{\rm conv}$ is the maximum timescale of the solar surface convection \citep{Hirzberger_1999_ApJ, Cranmer_2005_ApJ}. Approximating $F_{A,\ast}$ using the formula $F_{A,\ast} \approx \rho_\ast v_{A,\ast} \delta v_\ast^2$ (where $v_{A,\ast}$ denotes the Alfv\'en speed in the photosphere and $\delta v_\ast$ is the velocity perturbation), $\delta v_\ast = 1.2 {\rm \ km \ s^{-1}}$, which is comparable to the horizontal velocities observed in the photosphere \citep{Berger_1998_ApJ, Matsumoto_2010_ApJ_horizontal_velocity, Chitta_2012_ApJ, Oba_2020_ApJ}. While in reality, the presence of magnetic fields slightly alters the photosphere parameters \citep{Katsukawa_2005_ApJ}, we disregard this effect for simplicity.

While the large-scale structure of stellar magnetic field can be inferred using Zeeman Doppler imaging \citep{Semel_1989_AA, Brown_1991_AA, Donati_1997_AA, Donati_2006_MNRAS, Morgen_2012_AA, See_2019_ApJ}, the mid- to small-scale magnetic fields remain unobservable for stars other than the Sun. Due to the lack of sufficient information on the magnetic field structure, we simply assume that the stellar surface is divided into regions with and without magnetic fields, with the field strength ($B_\ast$) being uniform in the magnetic regions. The relationship between the unsigned magnetic flux and the filling factor ($f_\ast$) of the magnetic region is given by:
\begin{align}
    \Phi_\ast = 4 \pi R_\ast^2 B_\ast f_\ast. \label{eq:surface_magnetic_flux}
\end{align}

Magnetic fields on the solar surface are localized, with a flux density of approximately 1 kG \citep{Sanchez_Almeida_2000_ApJ, Dominguez_Cerdena_2006_ApJ, Tsuneta_2008_ApJ, Keys_2019_MNRAS, Van_Kooten_2024_arXiv}. This is nearly the equipartition field, where magnetic pressure equals gas pressure. Zeeman broadening observations of solar-type stars also suggest that the value of $B_\ast$ is on par with the equipartition field \citep{Saar_1996_proceedings, Cranmer_2011_ApJ}. For these reasons, in this study, we adopt values close to the equipartition field in the photosphere as follows.
\begin{align}
    B_\ast = 1340 {\rm \ G}.
\end{align}
However, it is noteworthy that in structures with intense magnetic fields, such as sunspots, the field strength exceeds the equipartition value \citep{Livingston_2006_SoPh, Okamoto_2018_ApJ, Castellanos_Duran_2020_ApJ}. This research consistently focuses on systems where energy is supplied by convective motions disturbing slender magnetic flux tubes, disregarding the effects of such large-scale intense magnetic structures.

\subsection{Basic equations \label{sec:basic_equations}}

In this study, we model magnetic loops (bundles of magnetic field lines forming loop structures) in stellar atmospheres. For a reduced computational cost, we adopt a one-dimensional magnetohydrodynamic system employed in the literature \citep{Moriyasu_2004_ApJ, Antolin_2010_ApJ, Washinoue_2019_ApJ, Washinoue_2021_MNRAS, Washinoue_2022_ApJ, Washinoue_2023_ApJ, Matsumoto_2023_arXiv} using the coordinate along the loop axis ($s$) and considering only spatial variations in $s$. The fundamental equations are then specified as follows (see Appendix A in \cite{Shoda_2021_AA} for derivation).
\begin{align}
    \frac{\partial}{\partial t} \vec{U} + \frac{1}{r^2 f} \frac{\partial}{\partial s} \left( \vec{F} r^2 f \right) = \vec{S},
    \label{eq:basic_equation_conservation_form}
\end{align}
where $r$ is the radial distance from the center of the star, $f$ is the filling factor of the flux tube, and 
\begin{align}
    \vec{U} =
    \left(
    \begin{array}{c}
    \rho \\
    \rho v_s \\
    \rho v_x \\
    \rho v_y \\
    B_x \\
    B_y \\
    e
    \end{array}
    \right), \hspace{1em}
    \vec{F} =
    \left(
    \begin{array}{c}
    \rho v_s \\
    \rho v_s^2 + p_T \\
    \rho v_s v_x - \dfrac{B_s B_x}{4\pi} \\
    \rho v_s v_y - \dfrac{B_s B_y}{4\pi} \\
    v_s B_x - v_x B_s \\
    v_s B_y - v_y B_s \\
    \left( e + p_T \right) v_s - \dfrac{B_s}{4 \pi} \left(\vec{v}_\perp \cdot \vec{B}_\perp \right) + q_{\rm cnd}
    \end{array}
    \right), \label{eq:conserved_variable_flux}
\end{align}
\begin{align}
    \vec{S}
    =
    \left(
    \begin{array}{c}
    0 \\
    \dfrac{1}{L}\left( p +\dfrac{1}{2} \rho \vec{v}_\perp^2 \right) - \rho \dfrac{GM_\ast}{r^2} \dfrac{dr}{ds}  \\
    \dfrac{1}{2L} \left( - \rho v_s v_x + \dfrac{B_s B_x}{4\pi} \right)  + \rho D^v_x \\
    \dfrac{1}{2L} \left( - \rho v_s v_y + \dfrac{B_s B_y}{4\pi} \right)  + \rho D^v_y \\
    \dfrac{1}{2L} \left( v_s B_x - v_x B_s \right)  + \sqrt{4 \pi \rho} D^b_x\\
    \dfrac{1}{2L} \left( v_s B_y - v_y B_s \right)  + \sqrt{4 \pi \rho} D^b_y\\
    - \rho v_s \dfrac{GM_\ast}{r^2} \dfrac{dr}{ds} - Q_{\rm rad}
    \end{array} 
    \right), 
    \label{eq:source_terms}
\end{align}
\begin{align}
    L = \left[ \frac{d}{ds} \ln \left( r^2 f \right) \right]^{-1},
\end{align}
\begin{align}
    &\vec{v}_\perp = v_x \vec{e}_x + v_y \vec{e}_y, \hspace{2em}
    \vec{B}_\perp = B_x \vec{e}_x + B_y \vec{e}_y, \\
    &p_T = p + \frac{\vec{B}_\perp^2}{8\pi}, \hspace{2em} e = e_{\rm int} + \frac{1}{2} \rho \vec{v}^2 + \frac{\vec{B}_\perp^2}{8\pi}.
\end{align}
Here, $x$ and $y$ denote the components transverse to the $s$ axis. $e_{\rm int}$ denotes the internal energy per unit volume and is approximated by that of partially ionized hydrogen gas (see Section 2.4 of Paper I for details). $q_{\rm cnd}$ stands for the conductive flux and is approximated by the Spitzer-H\"arm type formulation \citep{Spitzer_1953_PhysRev}:
\begin{align}
    q_{\rm cnd} = -\kappa_{\rm SH} T^{5/2} \frac{\partial T}{\partial s},
\end{align}
where $\kappa_{\rm SH} = 10^{-6}$ in the cgs unit.

\begin{figure}[t!]
\centering
\includegraphics[width=80mm]{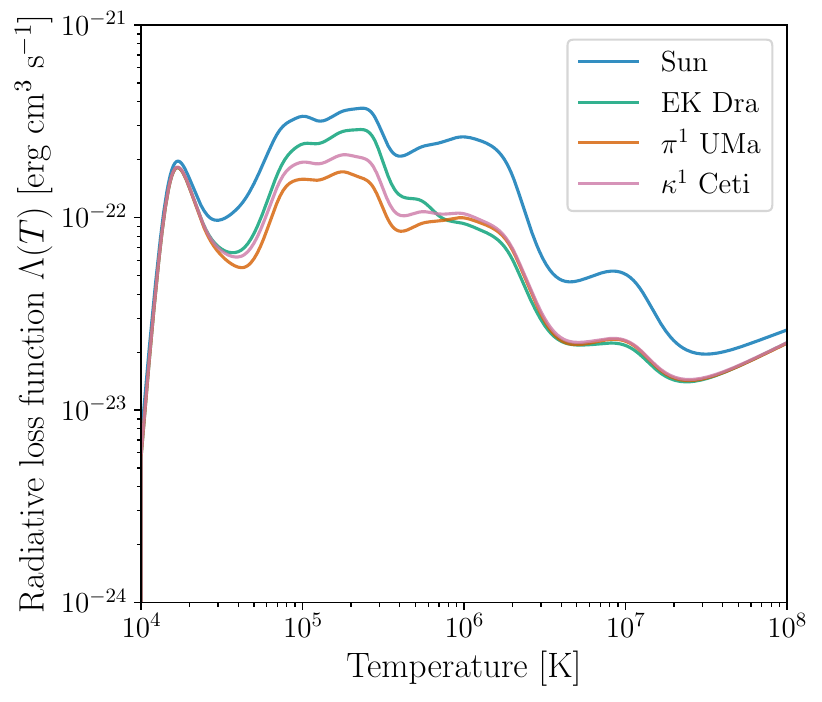}
\caption{
Comparison of radiative loss functions for four targets in this study.
}
\label{fig:Chianti_loss_function_comparison}
\end{figure}

$Q_{\rm rad}$ is the cooling rate by radiation and is given as follows (see Section 2.5 in Paper I for detail).
\begin{align}
    &Q_{\rm rad} = \xi_{\rm rad} Q_{\rm rad}^{\rm thck} + \left( 1 - \xi_{\rm rad} \right) Q_{\rm rad}^{\rm thin}, \\
    &Q_{\rm rad}^{\rm thck} = \frac{1}{\tau} \left( e_{\rm int} - e_{\rm int}^{\rm ref} \right), \hspace{2em} \tau = 0.1 {\rm \ s} \left( \frac{\rho}{\rho_\ast} \right)^{-1/2}, \\
    &Q_{\rm rad}^{\rm thin} = n_e n_{\rm H} \Lambda (T) \exp \left( -\frac{T_{\rm chr}^2}{T^2} \right), \label{eq:radiation_thin}\\
    &\xi_{\rm rad} = 1-  \exp \left( - \frac{10p}{p_\ast}\right),
\end{align}
where $n_e$ and $n_{\rm H}$ are the number densities of the electron and hydrogen atom, respectively. We set $T_{\rm chr} = 1.5 \times 10^4 {\rm \ K}$, which is slightly smaller than the value used in Paper I. $\Lambda (T)$ is the radiative loss function, whose functional form is given once the elemental abundance is prescribed. For the solar model, the reference abundance is provided by \citet{Schmelz_2012_ApJ}. The relationship between XUV spectrum and abundance is discussed in Section~\ref{sec:dependence_on_abundance}. For the stellar models, the coronal abundances constrained by X-ray observation \citep[][deduced by M1 method using MEKAL database]{Telleschi_2005_ApJ} are employed. Fig.~\ref{fig:Chianti_loss_function_comparison} compares $\Lambda (T)$ for the four targets in this study.

Numerous theoretical and observational studies have demonstrated that coronal heating is multi-dimensional in nature \citep{Parker_1972_ApJ, Einaudi_1996_ApJ, Rappazzo_2007_ApJ, Rappazzo_2008_ApJ, van_Ballegooijen_2011_ApJ, Cirtain_2013_Nature, Antolin_2014_ApJ, van_Ballegooijen_2017_ApJ_coronal_loop, Pagano_2020_AA, Antolin_2021_NatAs, Bose_2024_NatAs}. Consequently, one-dimensional models, unless modified, fail to accurately address coronal heating. Hence, we phenomenologically incorporate the multi-dimensional effect (turbulent cascade) in the corona through terms $D^v_{x,y}$ and $D^b_{x,y}$, which can be represented as follows \citep[][see also Section 2.6 in Paper I for detail]{Shoda_2018_ApJ_a_self-consistent_model}.
\begin{align}
    &D^v_{x,y} = - \frac{c_d}{4\lambda_\perp} \left( \left| z_{x,y}^+ \right| z_{x,y}^- + \left| z_{x,y}^- \right| z_{x,y}^+  \right), \label{eq:phenomenological_awt_vsource} \\ 
    &D^b_{x,y} = - \frac{c_d}{4\lambda_\perp} \left( \left| z_{x,y}^+ \right| z_{x,y}^- - \left| z_{x,y}^- \right| z_{x,y}^+  \right), \label{eq:phenomenological_awt_bsource}
    \\
    &\lambda_\perp = \lambda_{\perp,\ast} \frac{r}{R_\ast} \sqrt{\frac{f}{f_\ast}}, \label{eq:lambda_perp_via_f}
\end{align}
where $z_{x,y}^\pm = v_{x,y} \mp B_{x,y}/\sqrt{4 \pi \rho}$ and $\lambda_\perp$ denotes the perpendicular correlation length \citep{Abramenko_2013_ApJ, Sharma_2023_NatAs}. Following Paper I, we set $c_d=0.1$ \citep{van_Ballegooijen_2016_ApJ, Verdini_2019_SolPhys}.

\subsection{Loop characteristics}
In Paper I, a vertically extended magnetic field structure was assumed to simplify the calculation of the DEM. However, actual coronal loops are better represented as semi-circular \citep{Aschwanden_2002_SolPhys, Reale_2014_LRSP}. Therefore, we consider a semi-circular loop structure, unlike in Paper I, realized by establishing a specific relationship between $r$ and $s$ as follows.
\begin{align}
    r = R_\ast + \frac{2 L_{\rm half}}{\pi} \sin \left( \frac{\pi s}{2 L_{\rm half}} \right),
\end{align}
where $L_{\rm half}$ is the half loop length. 

The expansion of coronal loops is prescribed in terms of the filling factor of the flux tube ($f$). We note that the cross-sectional area of the flux tube is expressed as $r^2 f$. In this work, $f$ is specified as a function of the radial distance as follows.
\begin{align}
    f = \frac{e^{\left( r-R_\ast \right)/(2H_\ast)}}{\sqrt{e^{\left( r-R_\ast \right)/H_\ast} + f_\ast^{-2} -1}},
\end{align}
where $f_\ast$ denotes the filling factor at the photosphere (given by Eq.~\eqref{eq:surface_magnetic_flux}) and $H_\ast$ represents the pressure scale height at the photosphere. We note that the following asymptotic properties are satisfied:
\begin{align}
    f_{r \approx R_\ast} \approx f_\ast e^{\left( r-R_\ast \right)/(2H_\ast)}, \ \ \ \ \lim_{r \to \infty} f = 1.
\end{align}
Given that $H_\ast$ represents the pressure scale height, the first equation implies that the magnetic flux tube expands with a constant plasma beta. From the magnetic flux conservation, the magnetic field strength along the loop axis, $B_s$, is given by:
\begin{align}
    B_s = B_\ast R_\ast ^2f_\ast/(r^2 f). \label{eq:axial_magnetic_field_via_f}
\end{align}

According to this formulation, the coronal magnetic field strength is represented approximately by $B_\ast f_\ast$. For the solar minimum and maximum, the coronal field strengths are approximately 5.4 G and 9.0 G, respectively (see Table~\ref{table:target_parameters}). These values fall within the observed range of field strengths for warm coronal loops \citep{Nakariakov_2001_AA, Verwichte_2004_SolPhys} but are lower than the typical field strengths of active regions \citep{Lin_2000_ApJ, Jess_2016_NatPhys}. The applied field length, along with $f=1$ in the corona, suggests that our model presumes the corona to be filled with warm loops.

One might argue that adding the contribution of the active region would result in a higher model output. In other words, our model likely represents the lower limit of XUV emission as it does not consider the active regions, in particular the hot loops in the core of the active regions \citep{Del_Zanna_2015_AA, Barnes_2019_ApJ}. However, because the surface magnetic flux is fixed, adding the active region reduces the magnetic flux of the ambient corona, which may decrease its XUV emission. Therefore, it is uncertain whether including the active region under fixed surface magnetic flux would enhance the total XUV emission. The effect of variations in coronal field strength needs further investigation.

\subsection{Boundary condition}
In this model, both boundaries are posited on the stellar surface, with $X_{\rm L}$ and $X_{\rm R}$ denoting the value of a physical quantity $X$ at the left ($s=0$) and right ($s=2L_{\rm half}$) boundaries, respectively.

In our model, atmospheric heating energy is supplied by waves injected from the boundary (photosphere). For simplicity, only Alfv\'en waves are considered \citep[but see also][for the possible contribution of longitudinal waves]{Morton_2019_Nature_Astronomy, Shimizu_2022_ApJ, Kuniyoshi_2024_ApJ_pmode_conversion, Huang_2024_arXiv}. Since Alfv\'en waves are incompressible transverse waves, they do not cause disturbances in density, temperature, or longitudinal velocity in the linear regime. Therefore, the fixed boundary conditions are imposed on density, temperature, and longitudinal velocity as follows.
\begin{align}
    \rho_{\rm L} = \rho_{\rm R} = \rho_\ast, \ \ \ T_{\rm L} = T_{\rm R} = T_\ast, \ \ \ v_{s,{\rm L}} = v_{s,{\rm R}} = 0.
\end{align}
Given that the footpoints of coronal loop reside on the photosphere, the filling factor is expressed as
\begin{align}
    f_{\rm L} = f_{\rm R} = f_\ast
\end{align}

The boundary conditions of the transverse components are expressed using Els\"asser variables \citep[$\vec{z}^\pm_\perp = \vec{v}_\perp \mp \vec{B}/\sqrt{4 \pi \rho}$, ][]{Elsasser_1950_PR, Marsch_1987_JGR, Magyar_2019_ApJ}. It is important to note that in our computational setup, $\vec{z}^+_\perp$ and $\vec{z}^-_\perp$ represent the amplitudes of Alfv\'en waves propagating in the same and opposite directions as the $s$-axis, respectively. Alfv\'en waves are reflected in the chromosphere and transition region \citep{Cranmer_2005_ApJ, Verdini_2007_ApJ, Reville_2018_ApJ, Murabito_2024_PhRvL}, and the reflected waves leave the computational domain through the lower boundary due to the free boundary condition.
\begin{align}
    \left. \frac{\partial}{\partial s} \vec{z}^-_\perp \right|_{\rm L} = \left. \frac{\partial}{\partial s} \vec{z}^+_\perp \right|_{\rm R} = 0.
\end{align}
In Paper I, the amplitude of upward waves propagating from the photosphere was fixed as an input parameter. However, due to the interaction of surface thermal convection and magnetic fields, maintaining a constant amplitude of specific waves seems somewhat unnatural. Instead, we assume that the net Poynting flux propagating from the surface upwards remains constant, i.e.,
\begin{align}
    \overline{F_{A,{\rm L}}} = - \overline{F_{A,{\rm R}}} = F_{A,\ast},
\end{align}
where
\begin{align}
    F_A = \frac{1}{4} \rho v_A \left( {\vec{z}_\perp^+}^2 - {\vec{z}_\perp^-}^2 \right), \ \ \ v_A = \frac{B_s}{\sqrt{4 \pi \rho}},
\end{align}
and the overline indicates the average over a period three times the maximum wave period.

The temporal dependence of Els\"asser variables is set such that the power spectrum with respect to frequency ($\omega$) is described by $1/\omega$ \citep{Van_Kooten_2017_ApJ}, that is,
\begin{align}
    \vec{z}_{\perp, {\rm L}}^+, \ \vec{z}_{\perp, {\rm R}}^- \propto \int_{2\pi/\tau^{\rm max}}^{2\pi/\tau^{\rm min}} \omega^{-1/2} \sin \left[ \omega t + \phi(\omega) \right] d\omega,
\end{align}
where $\tau^{\rm max}$ and $\tau^{\rm min}$ correspond to the maximum and minimum periods of the waves, respectively, and $\phi(\omega)$ is a random phase function ranging in $0 \le \phi (\omega) < 2 \pi$.
The maximum period is determined by the turnover time of convection, expressed as 
\begin{align}
    \tau^{\rm max} = 1/f^{\rm conv}_\ast = 1000 {\rm \ s}.
\end{align}
Although determining the minimum period in the solar atmosphere is challenging, considering the significant energy of waves with periods of several tens of seconds \citep{He_2009_AA, Okamoto_2011_ApJ, Srivastava_2017_ScientificReports, Petrova_2023_ApJ, Lim_2023_ApJ, Lim_2024_arXiv}, we set
\begin{align}
    \tau^{\rm min} = \tau^{\rm max}/100 = 10 {\rm \ s}.
\end{align}

\subsection{Grid setup and numerical solver \label{sec:model:grid}}

In the stellar atmosphere, the scale length varies by 2-3 orders of magnitude due to temperature inhomogeneity. This scale difference results in the required grid size varying in space. To avoid using an excessively high number of grids, we define a non-uniform grid as follows. Let the total number of grids be $N$, the centre  of the $i$-th grid ($0 \le i \le N-1$) be $s_i$, and its width be $\Delta s_i$. For $1 \le i < N/2-1$, we iteratively define $s_i$ and $\Delta s_i$ as follows.
\begin{align}
    &\Delta s_i = \max \left[ \Delta s_{\rm min}, \min \left[ \Delta s_{\rm max}, \Delta s_{\rm min} + \frac{2\varepsilon_{\rm ge}}{2 + \varepsilon_{\rm ge}} \left( s_{i-1} - s_{\rm ge} \right) \right] \right], \nonumber \\
    &s_i = s_{i-1} + \frac{1}{2} \left( \Delta s_{i-1} + \Delta s_i \right),
\end{align}
where $s_0 = 0$, $\Delta s_0 = \Delta s_{\rm min} = 1 {\rm \ km}$, $\Delta s_{\rm max} = 500 {\rm \ km}$. In modeling the solar activity minimum and maximum, $s_{\rm ge}$ is set to $2 \times 10^4$ km, while for other cases, it is set to $5 \times 10^3$ km. This differentiation is due to the different transition region heights between the Sun and other stars. We have confirmed that in all cases, the transition region remains below $s_{\rm ge}$, ensuring it is resolved at a resolution of $\Delta s_{\rm min}$. The parameter $\varepsilon$ was adjusted to values between 0.01 and 0.02. For $i \ge N/2$, grid widths are set symmetrically around the centre of the computational domain,
\begin{align}
    \Delta s_i = \Delta s_{N-1-i}, \ \ \ \ s_i = s_{i-1} + \frac{1}{2} \left( \Delta s_{i-1} + \Delta s_i \right).
\end{align}

To solve the basic equations numerically, we rewrite them using the cross-section-weighted variables (see Section 2.8 in Paper I). This allows the application of the finite volume method as in the Cartesian coordinate. We use the HLLD approximated Riemann solver \citep{Miyoshi_2005_JCP} for computing numerical flux at cell boundaries. A fifth-order monotonicity-preserving reconstruction \citep{Suresh_1997_JCP} is used in the uniform-grid regions and the second-order MUSCL reconstruction \citep{van_Leer_1979_JCP} in the non-uniform-grid regions. Time integration is performed using the 3rd-order SSP Runge-Kutta method \citep{Shu_1988_JCP, Gottlieb_2001_SIAMR}. Thermal conduction is solved separately using the second-order super-time-stepping method \citep{Meyer_2012_MNRAS, Meyer_2014_JCP}.

\subsection{DEM and spectrum calculation}

The calculation of the XUV spectrum is conducted as a post-process, adhering to the following procedure. First, we assume that all XUV emissions are optically thin, a generalization that is not universally accurate \citep{Schrijver_1994_AA, Warren_1998_JGR, Avrett_2008_ApJ} but serves as a reasonable approximation across the XUV range. These emissions are calculated based on the time-averaged DEM, defined in Eq.~\eqref{eq:DEM_definition}, reflecting our interest in the steady component of XUV emission. As in Paper I, the contribution function $G_\lambda (T)$ for each wavelength is computed using CHIANTI version 10 \citep{Dere_1997_AA, Del_Zanna_2021_ApJ}, and the intensity at each wavelength $I_\lambda$ is determined using Eq.~\eqref{eq:intensity_dem}. The abundance used in calculating the contribution function aligns with that employed in computing the radiative loss function (see Section~\ref{sec:basic_equations}). These processes are carried out using the open-source package ChiantiPy. Finally, assuming uniform brightness in the XUV band, the flux at distance $r$ is calculated using the formula \citep{Rybicki_1979_book}:
\begin{align}
    F_\lambda = \pi I_\lambda \left( \frac{R_\ast}{r} \right)^2.
\end{align}

\section{Validation with solar observations}

In Paper I, we demonstrated that the one-dimensional coronal heating model has the potential to reproduce the solar XUV spectrum. However, in Paper I, we only compared the calculation results for specific parameters with a specific observation, and thus, the investigation into the limitations and uncertainties of the model was insufficient. Without understanding these aspects, we are not ready for meaningful comparisons with stellar observations. Therefore, in this section, we aim to understand the characteristics of the model output through a more detailed comparison with solar observations.

The model validation in Paper I was insufficient in three respects.
First, while Paper I conducted model validations for a solar activity minimum, it is critical to also validate the model for a solar activity maximum to better assess its applicability, particularly for flaring stars like EK Dra \citep{Audard_2000_ApJ}. Second, it assumed a constant coronal loop length across space and estimates the whole coronal emission from a single loop calculation. Given the diversity in coronal loop lengths \citep{Aschwanden_2002_SolPhys, Mac_Cormack_2022_AdSpR}, it is necessary to consider the distribution of loop lengths for a better comparison between model predictions and observations. Lastly, in Paper I, the elemental abundance was fixed to the typical coronal value \citep{Schmelz_2012_ApJ}. However, given that elemental abundance changes with space and time \citep{Del_Zanna_2023_ApJS}, accounting for the associated uncertainties in the model outputs is essential. We will address these issues in the following sections.

\begin{figure}[t!]
\centering
\includegraphics[width=75mm]{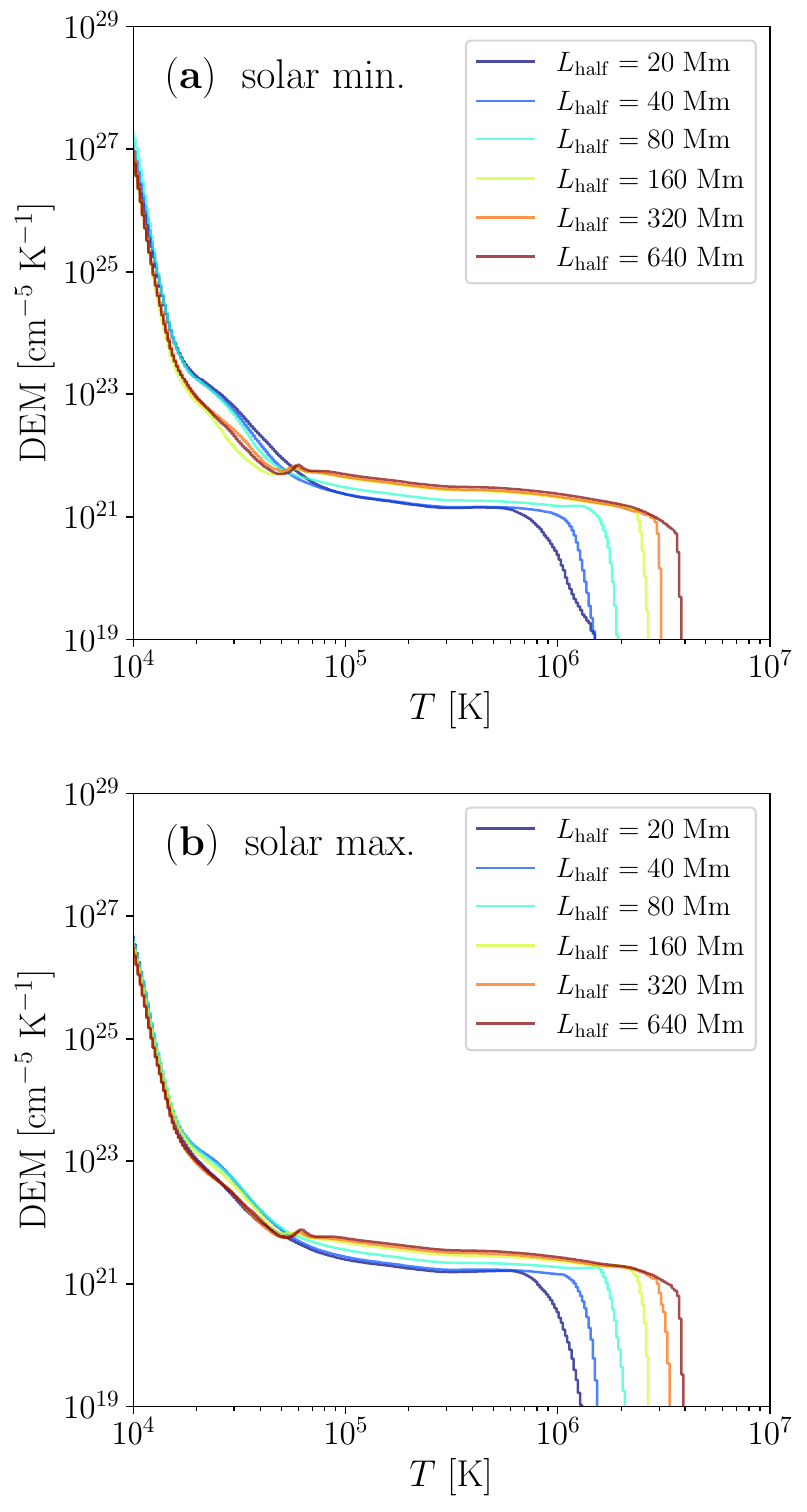}
\caption{Dependence of DEM on loop length. a: activity minimum. b: activity maximum.}
\label{fig:Loop_length_dependence_solar_DEM}
\end{figure}

\begin{figure*}[t!]
\centering
\includegraphics[width=150mm]{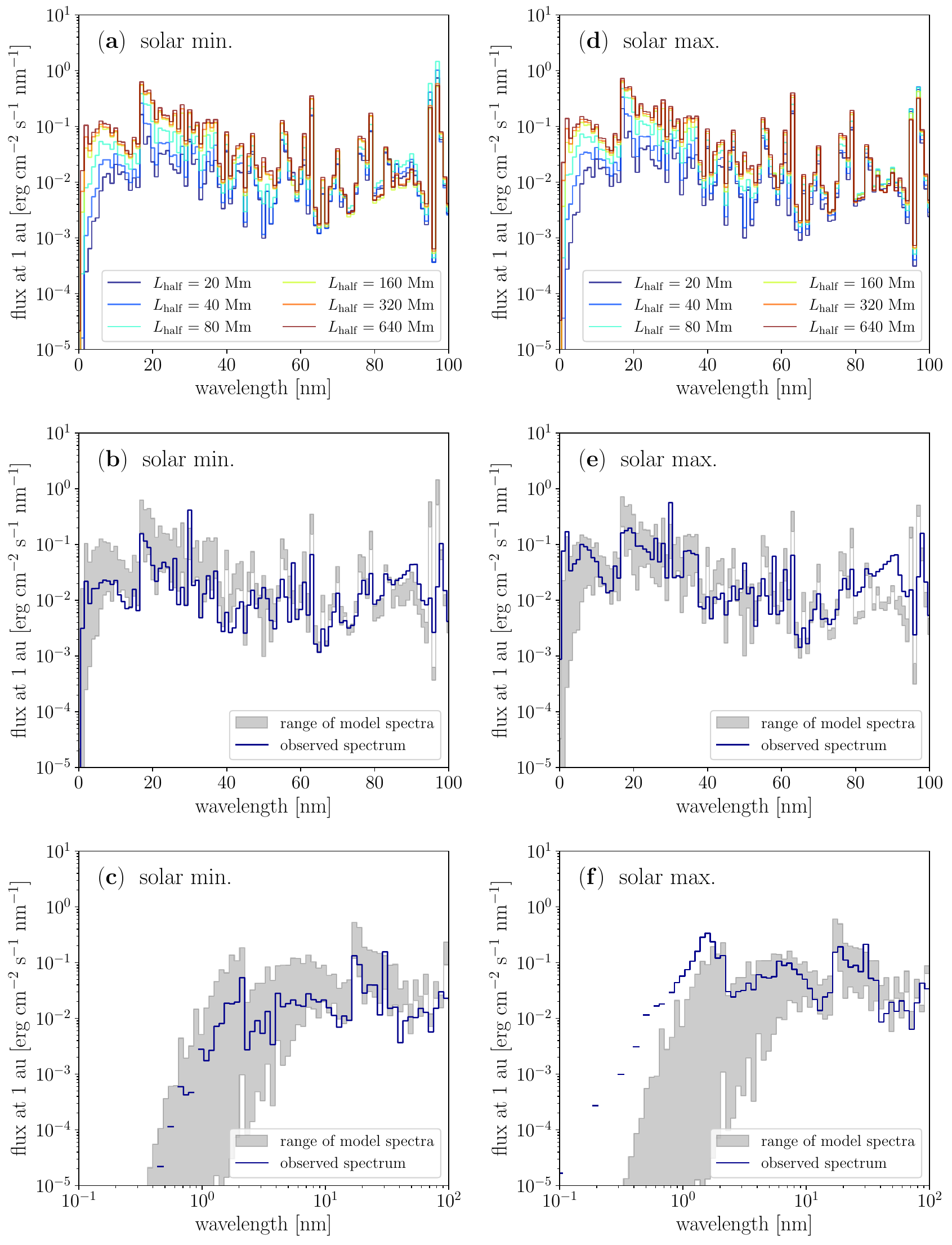}
\caption{Dependence of the XUV emission spectrum on loop length and comparison between model outputs and observations. The left panels (a, b, c) show data for solar minimum, while the right panels (d, e, f) display data for solar maximum. Top panels (a, d): Loop-length dependence of the XUV spectrum, with different colors representing different loop lengths. For clarity, the spectra are averaged over 1 nm intervals. Middle panels (b, e): Comparison of the model output range (gray shaded area) and observational data (blue solid line). Similar to the top panels, the spectra are averaged over 1 nm intervals. Bottom panels (c, f): Similar to the middle panels, but the spectra are averaged over bins that are evenly divided on a logarithmic scale (see text for details).}
\label{fig:loop_length_dependence_solar_vertical}
\end{figure*}

\subsection{Assessing the model spectra at solar minimum and maximum \label{sec:model_validation_minimum_maximum}}

To validate the model for both the solar activity minimum and maximum, simulations are conducted using the respective surface magnetic flux values (see Table~\ref{table:target_parameters}) as input parameters. For simplicity, the elemental abundance is fixed to be at the typical coronal value \citep{Schmelz_2012_ApJ} regardless of activity level. To account for the diversity of loop lengths, we conduct calculations for half-loop lengths ($L_{\rm half}$) of 20, 40, 80, 160, 320, and 640 Mm. The minimum  value of $L_{\rm half}$ (20 Mm) corresponds to the shortest loop length capable of reproducing coronal heating, whereas the maximum value (640 Mm) approximates the longest estimated loop length in the solar corona \citep{Mac_Cormack_2017_ApJ}.

Fig.~\ref{fig:Loop_length_dependence_solar_DEM} shows the loop-length dependence of differential emission measures (DEMs) for solar minimum (top) and maximum (bottom). The DEMs presented are time-averaged over the period $4.8 \times 10^5 {\rm \ s} < t < 9.6 \times 10^5 {\rm \ s}$. The averaging period is selected to ensure the system reached a quasi-steady state beforehand and exceeds the timescale of the variations in coronal density and temperature. As the loop length increases, the DEM extends to higher temperatures, consistent with analytical studies showing that longer loops, given the same energy flux, lead to higher coronal temperatures \citep{Rosner_1978_ApJ, Serio_1981_ApJ}. This trend is also supported by more realistic numerical simulations \citep{Bourdin_2016_AA, Shoda_2021_AA}. The coronal temperatures range from 0.78 to 3.72 MK for the solar minimum and from 0.85 to 3.78 MK for the solar maximum, in agreement with observational estimates \citep{Johnstone_2015_AA}.

The top panels of Fig.~\ref{fig:loop_length_dependence_solar_vertical} illustrate the loop-length dependence of the XUV spectra (calculated using the DEMs shown in Fig.~\ref{fig:Loop_length_dependence_solar_DEM}). The spectra are averaged over 1-nm bins for clarity. For both solar minimum and maximum, the dependence on loop length varies by wavelength range; for relatively long wavelengths ($\lambda > 40$ nm), the dependence is minor, while for shorter wavelengths ($\lambda < 40$ nm), we find a significant difference reaching up to an order of magnitude. This trend, as previously suggested in Paper I, is likely due to the different behavior of line formation temperature as a function of wavelength. Low-energy EUV lines are emitted from low-temperature ($10^{4-5} {\rm \ K}$) plasmas, while X-ray and high-energy EUV lines originate from high-temperature ($10^{6-5} {\rm \ K}$) plasmas \citep{Shoda_2021_AA}. Consequently, the emissivity of X-ray is more sensitive to the coronal maximum temperature (and the loop length).

Notably, for emission from the lower ($T \approx 10^4 {\rm \ K}$) transition region ($\lambda = 70-90$ nm), the dependence on loop length is opposite to that of X-rays, with radiation intensity decreasing as loop length increases. This occurs because the DEM in the lower transition region is reduced when the loop length is longer, as faintly observed in Fig.~\ref{fig:Loop_length_dependence_solar_DEM}. The trend of DEM decreasing at $T \approx 10^4 {\rm \ K}$ with increasing coronal temperature is more evident in the stellar simulations (Section~\ref{sec:comparion_with_stellar_observations}) and is likely caused by the thinning of the transition region due to a hotter corona and stronger thermal conduction \citep{Johnston_2021_AA_LTRAC, Iijima_2021_ApJ}.

In the middle and bottom panels of Fig.~\ref{fig:loop_length_dependence_solar_vertical}, the range of model spectra, shown as a gray-shaded area, is compared with the observed spectra (blue line). The middle panels compare spectra averaged over 1 nm intervals, while the bottom panels compare spectra averaged over bins set evenly on a logarithmic scale. Here, the bin size on the logarithmic scale is set to be $10\%$ of the wavelength; for example, at a wavelength of 30 nm, the bin size is 3 nm. 

For wavelengths below 20 nm, the XUV spectrum observed during the solar minimum aligns with the model spectra range. However, during the solar maximum, the high-energy X-ray spectrum ($\lambda < 1 {\rm \ nm}$) deviates significantly from the model range. We suspect this deviation is due to flaring activity, as it is not observed during the solar minimum. This suggests that our current model underestimates high-energy X-rays, especially those from flaring stars.

In the 20-60 nm range, while the averaged values over this wavelength band roughly agree between the model and observation, the observed values deviate from the model predictions at certain wavelengths. The discrepancy is partly caused by Helium emission, which makes up a notable fraction of the emission within this wavelength range. For instance, to replicate the He I continuum around 50.4 nm, it is necessary to solve the radiative transfer equation with considerations for photoionization by coronal illumination \citep{Zirin_1975_ApJ, Andretta_1997_ApJ, Andretta_2003_AA}. Additionally, the DEM-based method is known to underestimate He emission lines (25.6 nm, 30.4 nm, 58.4 nm), a well-known issue \citep[helium line enhancement,][]{Jordan_1975_MNRAS, MacPherson_1999_MNRAS, Judge_2003_ApJ, Pietarila_2004_ApJ} potentially resolvable by considering non-equilibrium ionization \citep{Golding_2017_AA}. Hence, our current model cannot reproduce the intensity of each emission line in the EUV band, though the wavelength-integrated flux is reasonably accurate.

In the 60-90 nm range, the Lyman continuum is underestimated during both maximum and minimum phases, with the difference being particularly notable during the maximum phase. Several factors could explain this discrepancy. First, since the solar atmosphere is not optically thin in the Lyman continuum \citep{Avrett_2008_ApJ, McLaughlin_2023_ApJ}, our model inherently cannot describe it accurately. Second, as the Lyman continuum in our model is emitted from the lower transition region ($T \approx 10^4 {\rm \ K}$), we may overestimate cooling or underestimate heating in this region. In fact, our model assumes a constant coronal backradiation effect (as represented by $T_{\rm chr}$ in Eq.~\eqref{eq:radiation_thin}) irrespective of activity levels, which might lead to underestimation of backradiation heating and the Lyman continuum emission in the solar maximum. Therefore, it should be noted that the accuracy of the Lyman continuum is not guaranteed in our model.

\begin{figure*}[t!]
\centering
\includegraphics[width=150mm]{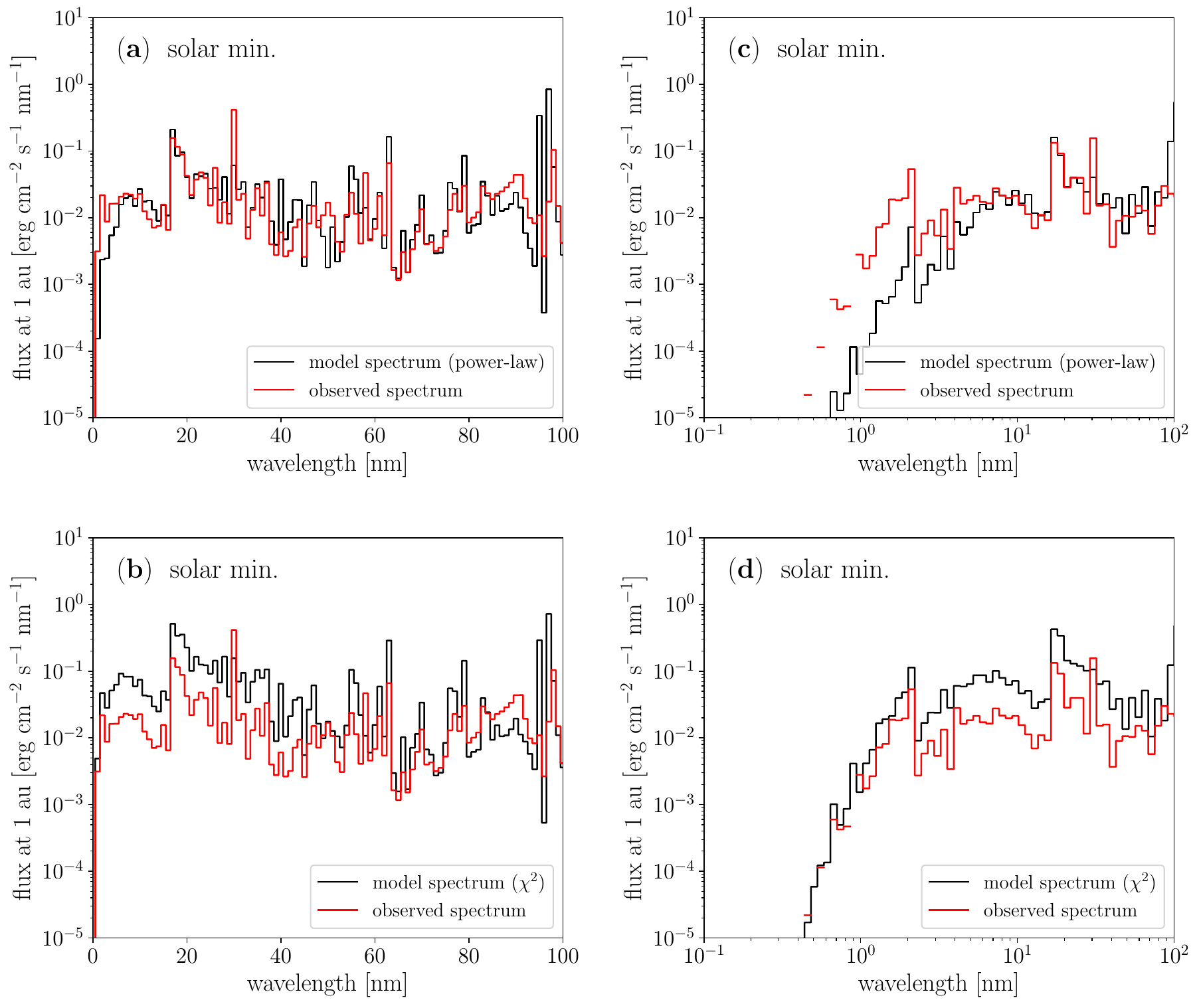}
\caption{Comparison between the model spectrum obtained by average over loop lengths (black) and the observed spectrum (red) for the solar activity minimum. The left side compares the averages in bins equally divided in linear scale, while the right side compares the averages in bins equally divided in logarithmic scale (bin size defined as in Fig.~\ref{fig:loop_length_dependence_solar_vertical}). The top panels display the averaged model spectrum weighted by the power-law distribution, and the bottom panels display the averaged model spectrum weighted by the chi-squared distribution.}
\label{fig:XUV_spectrum_superposition_comparison_solarmin}
\end{figure*}

\begin{figure*}[t!]
\centering
\includegraphics[width=150mm]{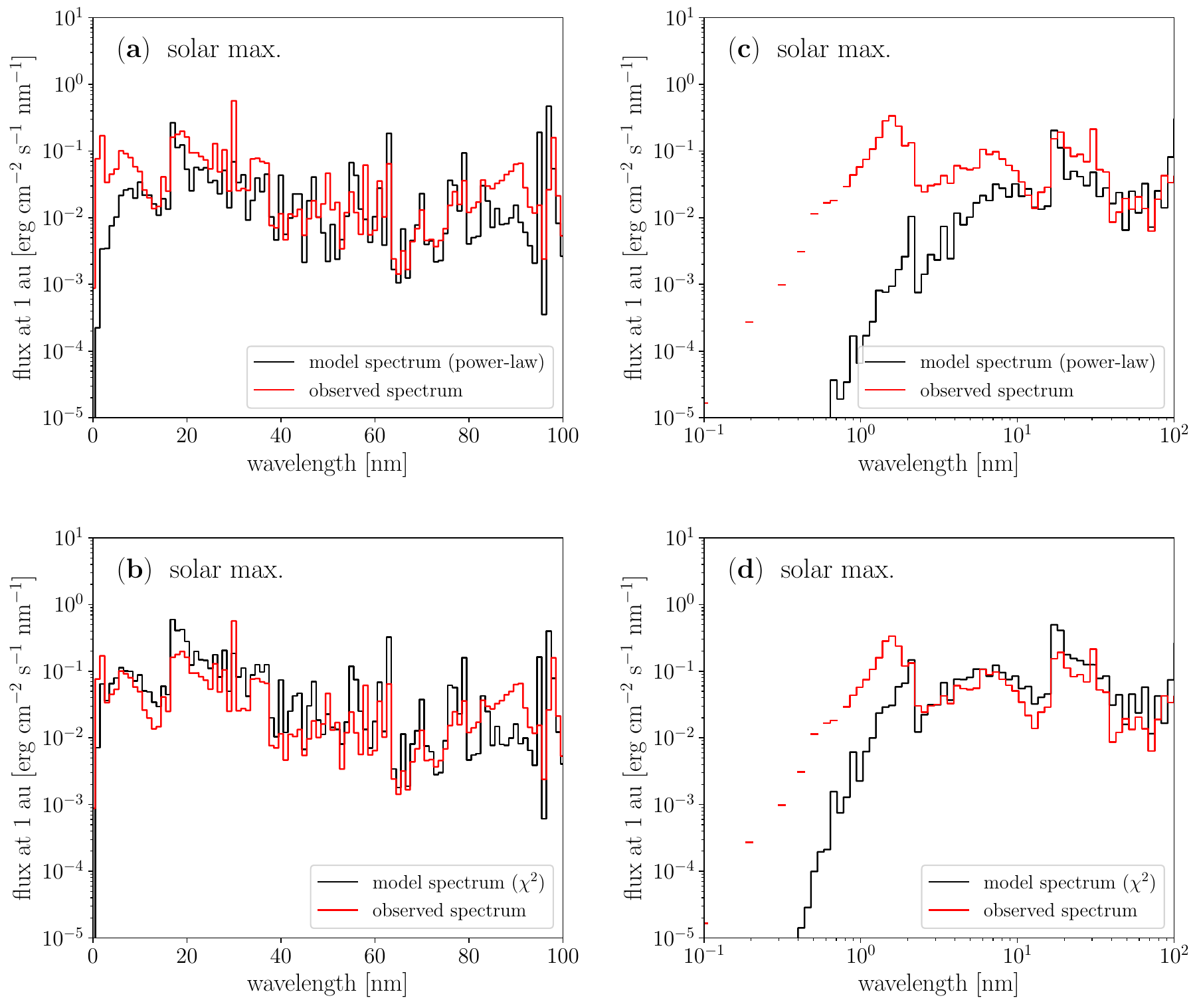}
\caption{Same as Fig.~\ref{fig:XUV_spectrum_superposition_comparison_solarmin} but for the solar activity maximum.}
\label{fig:XUV_spectrum_superposition_comparison_solarmax}
\end{figure*}

\begin{figure}[t!]
\centering
\includegraphics[width=69mm]{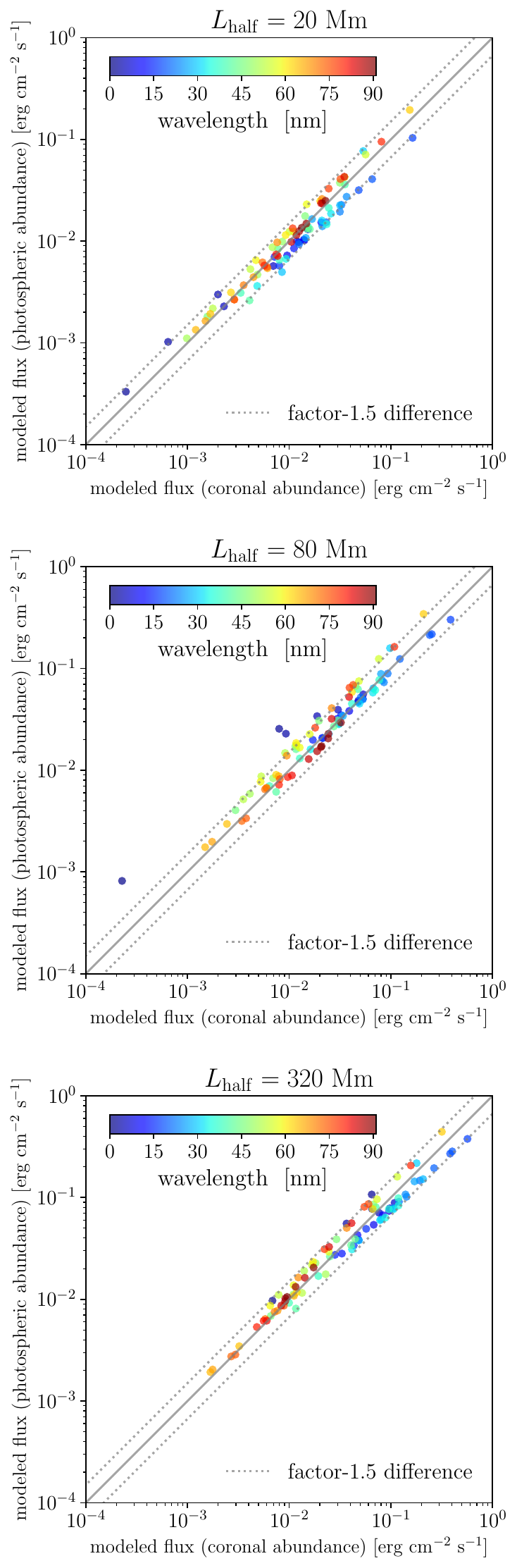}
\caption{Comparison of XUV radiation fluxes integrated per 1 nm between coronal- and photospheric-abundance cases. Symbol colors represent different wavelengths, and dashed lines indicate a 1.5-fold difference. Each panel corresponds to $L_{\rm half} = 20$, $80$, and $320$ Mm, respectively.}
\label{fig:XUV_spectrum_dependence_on_abundance}
\end{figure}

\subsection{Effect of loop-length variety}

To account for the effects of various coronal loop lengths, we calculate the weighted average of XUV spectra over multiple simulation runs with different loop lengths. This requires knowledge of the distribution of coronal loop lengths. However, to the authors' knowledge, no consensus has been reached regarding the loop-length distribution in the solar atmosphere. Therefore, we consider two proposed distributions from the literature to account for the effects of multiple lengths.

The frequency of occurrence of magnetic patches on the solar surface follows a power law with respect to magnetic flux \citep{Parnell_2009_ApJ, Iida_2012_ApJ}. Assuming the magnetic flux of a magnetic patch is proportional to its size, the distribution of the number of magnetic patches as a function of area $A$ also follows a power-law distribution with respect to $A$. Specifically, denoting the number of magnetic patches with areas between $A$ and $A+dA$ as $N(A)dA$, $N(A)$ satisfies the following relationship \citep{Takasao_2020_ApJ}:
\begin{align}
    N(A) \propto A^{-\alpha},
\end{align}
where $\alpha =1.85$. For simplicity, assuming that the coronal loop length is proportional to the square root of the size of the corresponding magnetic patches, the distribution with respect to the loop length can be expressed as follows:
\begin{align}
    N(L_{\rm half}) \propto L_{\rm half}^{-2\alpha + 1}. \label{eq:loop_length_distribution_power_law}
\end{align}

Another possible distribution of loop length is derived from an analysis that combines coronal magnetic field extrapolation and ultraviolet observations. This analysis indicates that the coronal loop length follows a chi-square-like distribution, with a mode at $L_{\rm half} = 100-150 {\rm \ Mm}$ \citep{Huang_2012_ApJ, Mac_Cormack_2020_AdSpR}. Therefore, we also consider the chi-square distribution given as follows.
\begin{align}
    N(L_{\rm half}) \propto L_{\rm half}^{k/2-1} \exp \left[ - \frac{(k-2)L_{\rm half}}{2L_{\rm mode}} \right], \label{eq:loop_length_distribution_chi_squared}
\end{align}
where $k$ is the degree of freedom and $L_{\rm mode}$ corresponds to the mode of this distribution. We set $k=4$ and $L_{\rm mode} = 120 {\rm \ Mm}$ to replicate the observed loop-length distribution.

Fig.~\ref{fig:XUV_spectrum_superposition_comparison_solarmin} compares the weighted averages of the model spectra with the observed spectrum in the solar minimum. The left and right panels illustrate the comparisons for the bin-averaged spectra on linear and logarithmic scales, respectively. The top panels show the power-law weighted average, based on Eq.~\eqref{eq:loop_length_distribution_power_law}, while the bottom panels present the chi-squared weighted average from Eq.~\eqref{eq:loop_length_distribution_chi_squared}. For the solar minimum, the power-law weighted average closely matches the observed data, whereas the chi-squared weighted average shows a greater difference. To better quantify the deviation between the model and observation, we calculate the mean error (ERR) as follows.
\begin{align}
    {\rm ERR} = \frac{1}{N_{\rm bin}} \sum_i \max \left( f_i^{\rm model}/f_i^{\rm obs}, f_i^{\rm obs}/f_i^{\rm model} \right) - 1,
\end{align}
where $N_{\rm bin}$ is the number of bins in the XUV range ($\lambda \le 91 {\rm \ nm}$) and $f^{\rm model/obs}_i$ represents the model/observed flux for the $i$-th bin. Using 1-nm bins to calculate ERR, we find ${\rm ERR} = 1.17$ for the power-law weighted average and ${\rm ERR} = 6.02$ for the chi-squared weighted average.

Fig.~\ref{fig:XUV_spectrum_superposition_comparison_solarmax} shows the comparison of model spectra and observation for the solar maximum with the same format as Fig.~\ref{fig:XUV_spectrum_superposition_comparison_solarmin}. In this case, the chi-squared weighted average aligns more closely with the observations compared to the power-law weighted average. Specifically, ${\rm ERR} = 2.26$ for the power-law weighted average and ${\rm ERR} = 1.49$ for the chi-squared weighted average. For both cases, we find the significant discrepancies in the high-energy X-ray region ($\lambda < 1 {\rm \ nm}$), as discussed in the previous section.

Summarizing the results on the superposition with respect to loop length, for both solar minimum and maximum, it is possible to generate model spectra that reproduce observations within a mean error of 2 by selecting an appropriate loop length distribution. Clearly, this does not fully assure the validity of our model. For a more sophisticated model validation, we need to derive the loop length distribution using magnetic field extrapolation from the observed magnetogram and compare the predicted and observed XUV spectra, which will be addressed in future work.

\subsection{Dependence of XUV emission on elemental abundance \label{sec:dependence_on_abundance}}

So far, we have calculated the radiative cooling function and the spectrum, assuming a fixed elemental abundance. The elemental abundance of the corona can slightly differ from that of the photosphere \citep{Pottasch_1963_ApJ}, and this discrepancy is suggested to depend on the first ionization potential (FIP) of each element, \citep{Feldman_1992_PhyS, Feldman_2003_SSRev, Lamming_2015_LRSP}. The relative abundance of elements in the corona compared to the photosphere is therefore referred to as the FIP bias \citep{von_Steiger_2000_JGR, Baker_2013_ApJ}.

The abundance used so far has been derived by setting the FIP bias for low-FIP elements to 2.14 \citep{Schmelz_2012_ApJ}. However, some observations suggest that the FIP bias for low-FIP elements in the quiet Sun is approximately 1, indicating an elemental abundance nearly identical to that of the photosphere \citep{Madsen_2019_ApJ, Del_Zanna_2023_ApJS}. To understand the uncertainty in the model spectrum due to the variation in abundance, we investigate how the XUV spectrum differs in the presence and absence of the FIP bias.

Fig.~\ref{fig:XUV_spectrum_dependence_on_abundance} compares the XUV spectra calculated with the typical coronal abundance \citep{Schmelz_2012_ApJ} and photospheric abundance \citep{Asplund_2009_ARAA}. The spectra are integrated per nanometer, converted to flux (erg cm$^{-2}$ s$^{-1}$), and compared for each wavelength within the XUV band ($\lambda \le 91 {\rm \ nm}$). The colors represent wavelengths, and the three panels show the comparison results for different loop lengths ($L_{\rm half} = 20, 80, 320 {\rm \ Mm}$). The solid line indicates where the two fluxes match, and the dashed lines represent a 1.5-fold difference.

It is evident that the abundance dependence of XUV emission is minimal regardless of loop length. This is likely because the increase in the radiative cooling function (due to metal enhancement) is offset by the decrease in coronal electron density, resulting in a weak dependence of XUV emission on abundance \citep{Washinoue_2023_ApJ}. The symbols in Fig.~\ref{fig:XUV_spectrum_dependence_on_abundance} are roughly within the factor-1.5 difference lines, and thus, the uncertainty in XUV emission due to abundance is estimated to be at most 50 $\%$.

\section{Model assessment with stellar XUV observations \label{sec:simulation_result}}

In this section, we aim to evaluate our model on young solar-type stars with significantly stronger magnetic activity than the Sun. For stars with limited observational data, we can only infer the lengths of specific coronal loops in indirect ways \citep{Reale_1998_AA, Jardine_2005_MNRAS, Lim_2022_ApJ}, and thus, calculating a weighted average as in solar models is not feasible. Therefore, we estimate the stellar XUV spectrum using a single-loop model and compare it with observations.

We assume that the typical loop length is largely independent of the level of stellar magnetic activity, based on the idea that it is determined by the scale at which the magnetic field is most energetic. For example, the typical loop length is a few times the stellar diameter when the dipolar field holds the maximum energy in the stellar magnetic field. Thus, the ratio of large-scale magnetic field strength ($\langle B_V \rangle$), observed via Zeeman-Doppler Imaging, to the total magnetic field strength ($\langle B_I \rangle$), observed via Zeeman Broadening, serves as an indicator of the typical loop length. A higher $\langle B_V \rangle/ \langle B_I \rangle$ ratio implies longer loops, as larger fraction of the magnetic energy is stored at the large scale. In the unsaturated phase, $\langle B_V \rangle$ and $\langle B_I \rangle$ follow similar scaling relations with the stellar Rossby number (Ro) \citep{Vidotto_2014_MNRAS, Reiners_2022_AA}:
\begin{align}
    \langle B_V \rangle \propto {\rm Ro}^{-1.38}, \ \ \ \langle B_I \rangle \propto {\rm Ro}^{-1.26},
\end{align}
indicating that the ratio $\langle B_V \rangle/\langle B_I \rangle$ remains nearly constant regardless of activity level. We set $L_{\rm half} = 80 {\rm \ Mm}$ as this is the most optimistic parameter for reproducing the solar XUV spectra both in the activity minimum and maximum.

According to the calculation results for solar activity minimum and maximum, the uncertainty in model output caused by fixing the half loop length at 80 Mm is approximately a factor of 3 across the entire XUV range. Additionally, the low-resolution calculations of $\pi^1$
UMa show that the error due to fixing the half loop length at 80 Mm is at most 0.86 for the X-ray luminosity and at most 3.12 for the XUV luminosity (see Appendix~\ref{appendix:loop_length_dependence_piuma} for details). Based on these considerations, we will compare the model output with observations, accounting for an uncertainty of a factor of 3.

\begin{figure}[t!]
\centering
\includegraphics[width=75mm]{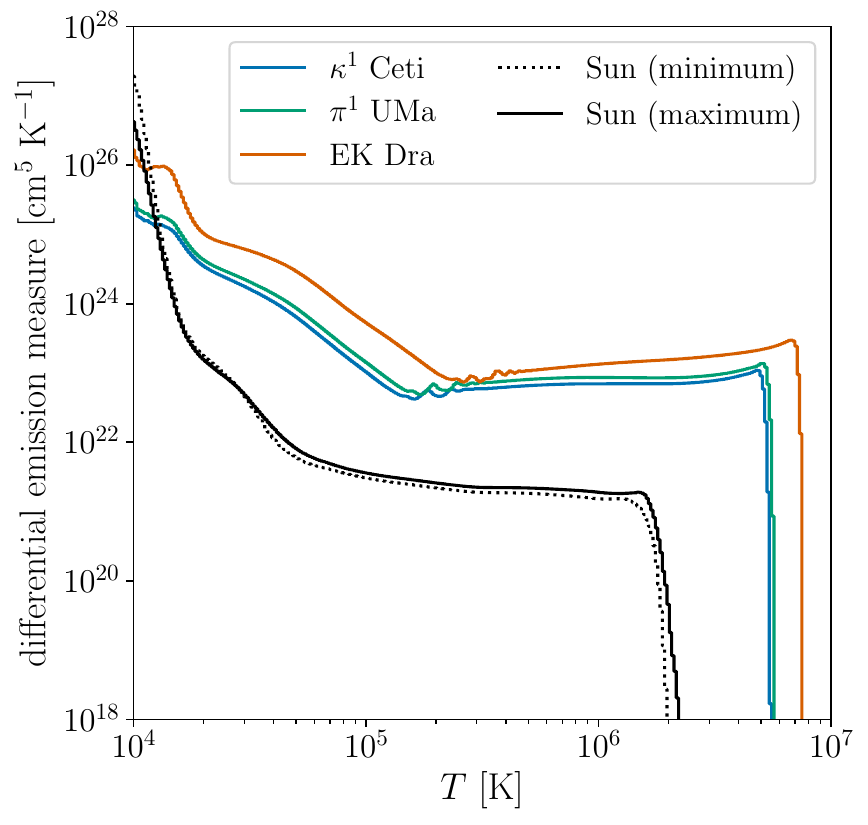}
\caption{Comparison of time-averaged differential emission measures (DEMs) calculated for the Sun, $\kappa^1$ Ceti, $\pi^1$ UMa, and EK Dra. The black lines represent the Sun (solid for activity maximum, dotted for activity minimum), while the colored lines correspond to the other three stars.}
\label{fig:DEM_comparison}
\end{figure}

\begin{figure*}[t!]
\centering
\includegraphics[width=150mm]{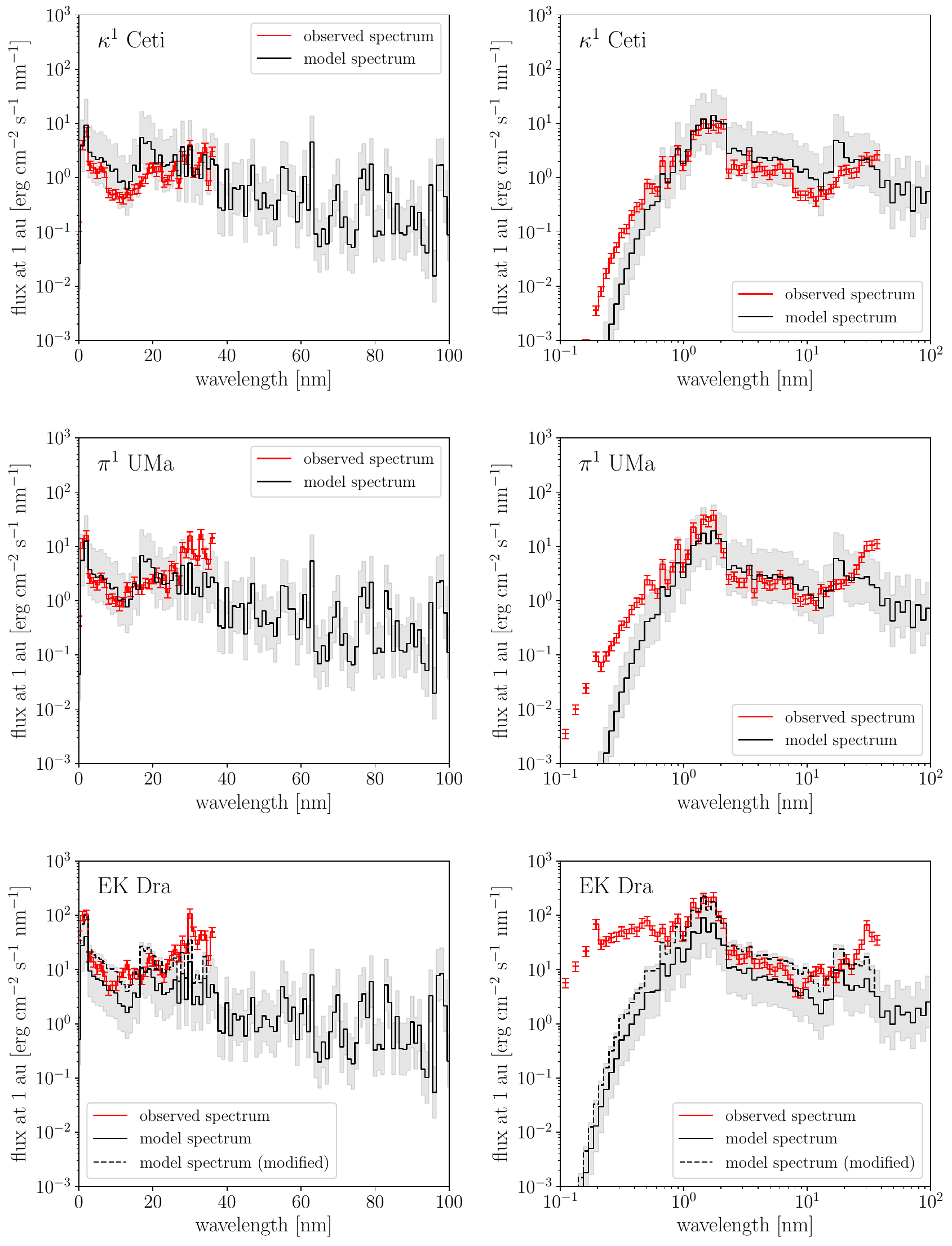}
\caption{The model spectra (black solid line) and observed spectra (red symbols) of young solar-type stars (top: $\kappa^1$ Ceti, middile: $\pi^1$ UMa, bottom: EK Dra) are compared. The left side shows the bin-averaged spectra on a linear scale, while the right side shows them on a logarithmic scale. The bin size definition follows Fig.~\ref{fig:loop_length_dependence_solar_vertical}. The gray area represents the uncertainty of the model (factor of 3). The black dashed line in the bottom panel indicates the spectrum after correction for the underestimation due to the insufficient resolution of the transition region (see the text for details).}
\label{fig:spectrum_bin_comaprison_solartype_stars}
\end{figure*}

\begin{figure*}[t!]
\centering
\includegraphics[width=150mm]{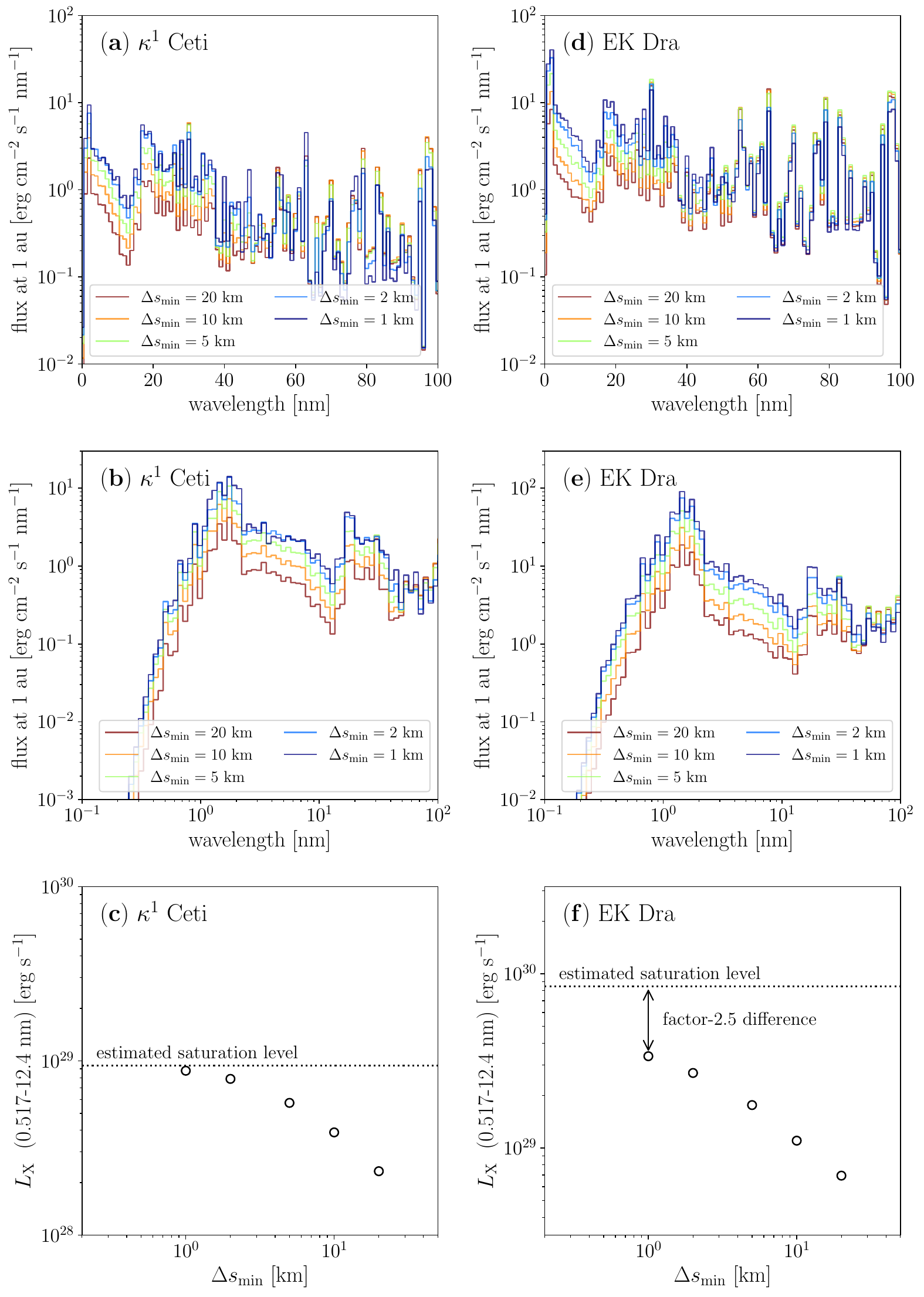}
\caption{The effect of transition region resolution ($\Delta s_{\rm min}$) on the XUV spectra of $\kappa^1$ Ceti (left) and EK Dra (right). The top and middle panels display the bin-averaged XUV spectra for varying resolutions, with different colors indicating different resolutions. The bin size definition follows Fig.~\ref{fig:loop_length_dependence_solar_vertical}. The bottom panels illustrate the computed X-ray luminosity as a function of $\Delta s_{\rm min}$. The dotted lines indicate the results obtained using the LTRAC method, where the transition region is artificially broadened. The X-ray luminosity is expected to saturate at this level when computed with adequate resolution.}
\label{fig:resolution_dependence}
\end{figure*}

\subsection{Comparison with the stellar XUV observations \label{sec:comparion_with_stellar_observations}}

Prior to comparing XUV spectra, it is beneficial to assess how variations in stellar parameters influence the DEM of the atmosphere. While the DEM and spectra are equivalent when the elemental abundance is known, the DEM is described directly by atmospheric physical characteristics, allowing for a clearer understanding of the fundamental physical differences.

Fig.~\ref{fig:DEM_comparison} compares time-averaged DEMs for the Sun (solar minimum and maximum), $\kappa^1$ Ceti, $\pi^1$ UMa, and EK Dra. Young solar-type stars possess significantly higher surface magnetic flux than the Sun, resulting in enhanced coronal heating. Specifically, the energy flux injected into the corona, as calculated by the model, amounts to $2.14 \times 10^7 {\rm \ erg \ cm^{-2} \ s^{-1}}$ for $\kappa^1$ Ceti, $2.54 \times 10^7 {\rm \ erg \ cm^{-2} \ s^{-1}}$ for $\pi^1$ UMa, and $7.20 \times 10^7 {\rm \ erg \ cm^{-2} \ s^{-1}}$ for EK Dra, markedly surpassing the solar values ($3.89\times 10^5 {\rm \ erg \ cm^{-2} \ s^{-1}}$ for the solar minimum, $4.61 \times 10^5 {\rm \ erg \ cm^{-2} \ s^{-1}}$ for the solar maximum). Consequently, the DEM extends to higher temperatures as the surface magnetic flux increases. In addition, the DEM becomes about two orders of magnitude larger than that of the Sun in $T>3 \times 10^4 {\rm \ K}$. This increase is due to the need for more plasma at the transition region to counteract the larger heat conduction from the corona through radiative cooling \citep{Rosner_1978_ApJ}. Interestingly, the DEM of the lower transition region ($T \approx 1.0 \times 10^4 {\rm \ K}$) is about one to two orders of magnitude lower than that of the Sun. A possible interpretation of this behavior, as discussed in Section~\ref{sec:model_validation_minimum_maximum}, is that an increase in coronal temperature reduces the thickness of the transition region \citep{Johnston_2021_AA_LTRAC, Iijima_2021_ApJ}, thereby decreasing the line-of-sight length contribution in DEM calculations.

We note that the DEM of the lower transition region may be underestimated in young solar-type stars. This is, as noted in Section~\ref{sec:model_validation_minimum_maximum}, because the DEM around $T \approx 1.0 \times 10^4 {\rm \ K}$ is sensitive to the effect of coronal backradiation, represented by the value of $T_{\rm chr}$. In the current model, $T_{\rm chr}$ is fixed, implying that the effect of backradiation remains constant regardless of the magnetic activity level. However, active stars should exhibit a greater backradiation effect, leading to increased heating of the chromosphere and lower transition region. To quantitatively assess this effect, a radiative transfer calculation is required, which remains a task for future research.

Fig.~\ref{fig:spectrum_bin_comaprison_solartype_stars} shows a comparison between the modeled and observed XUV spectra. The black solid lines represent the model spectra, while the red symbols indicate the observed spectra. \citet{Ribas_2005_ApJ} note that the observed XUV spectra may have errors up to $20 \%$ due to photon count noise and calibration inaccuracies. Therefore, we include error bars representing a $20 \%$ noise level in the observational data. It is important to note that $20 \%$ is an average value, and errors may be larger, especially in the wavelength range of 10-20 nm, where the signal is weak. The gray shaded areas indicate the uncertainty of the model.

For $\kappa^1$ Ceti and $\pi^1$ UMa, the model generally matches observations, supporting the accuracy of the model. In both stars, model values are significantly lower than observed values in $\lambda < 1 {\rm \ nm}$, similar to the solar maximum, indicating a significant contribution from flaring activity. In the case of EK Dra, the observed XUV spectrum systematically exceeds the model predictions. In particular in $\lambda < 1 {\rm \ nm}$, we find the model underestimates the observation by 1 to 3 orders of magnitude. The larger discrepancy in $\lambda < 1 {\rm \ nm}$ supports our hypothesis that flares contribute significantly to this band, considering the high frequency of large-scale flares in EK Dra \citep{Audard_2000_ApJ, Ayres_2015_AJ}.

The systematic underestimation of the model spectrum of EK Dra compared to observations may be attributed to insufficient resolution of the transition region. The extremely high coronal temperature of EK Dra (approximately $7 \times 10^6$ K) results in a very thin transition region. If the transition region is not resolved with sufficient precision, it may lead to an underestimation of the density above the transition region \citep{Bradshaw_2013_ApJ}, and consequently, a lower XUV spectrum than the actual one. To investigate this possibility, we conduct calculations for $\kappa^1$ Ceti and EK Dra with various resolutions at the transition region ($\Delta s_{\rm min}$), as shown in Fig.~\ref{fig:resolution_dependence}. We also perform simulations using the LTRAC method \citep[][see Appendix~\ref{appendix:ltrac} for details]{Iijima_2021_ApJ}, which allows us to avoid the transition-region problem and verify convergence with respect to resolution. Although the LTRAC method tends to overestimate the DEM at low temperatures, it accurately reproduces the DEM at high temperatures, thus allowing the calculation of X-ray luminosity close to the true value (see Appendix~\ref{appendix:ltrac}). The estimated saturation level is set to the X-ray luminosity obtained with the LTRAC method, and convergence is verified by checking if the X-ray luminosity reaches this level.

For $\kappa^1$ Ceti, the X-ray luminosity calculated with $\Delta s_{\rm min} = 1 {\rm \ km}$ is approximately the same as that calculated with the LTRAC method, indicating that the resolution convergence is ensured. We expect the similar result for $\pi^1$ UMa, where the magnetic flux is almost the same. However, for EK Dra, the X-ray luminosity calculated with $\Delta s_{\rm min} = 1 {\rm \ km}$ is only about $40 \%$ of that calculated with the LTRAC method, suggesting that convergence with respect to resolution has not yet been achieved.

As shown in Panel (d) of Fig.~\ref{fig:resolution_dependence}, the resolution dependence of the spectrum is noticeable for $\lambda < 35 {\rm \ nm}$, with the spectrum increasing almost uniformly across wavelengths in this range as the resolution increases. Considering this fact and the estimated true X-ray luminosity being 2.5 times that of the current model, we create a modified model spectrum by increasing the model spectrum by 2.5 times for wavelengths below 35 nm, as indicated by the dashed line in the bottom panel of Fig.~\ref{fig:spectrum_bin_comaprison_solartype_stars}. The modified model spectrum agrees with the observation, and the systematic gap between the model and observations is no longer observed. Therefore, the difference between the model and observations seen in EK Dra is likely due to insufficient resolution, and this gap is expected to be resolved with adequate resolution.

However, the observed spectrum of EK Dra in the 25-35 nm range remains significantly higher compared to the modified model spectrum (Figure~\ref{fig:spectrum_bin_comaprison_solartype_stars}). A similar trend is also observed in $\pi^1$ UMa. We must acknowledge the possibility that the current model may fail to accurately reproduce the spectrum within the 25-35 nm range, although the reason for this discrepancy is still unknown.

Overall, our model reproduces the observed XUV spectra of $\kappa^1$ Ceti, $\pi^1$ UMa, and possibly EK Dra, in the wavelength range of 1-30 nm, meaning that model-based estimation of the stellar XUV spectra is feasible for young solar-type stars with surface magnetic fluxes 100-300 times that of the Sun.

\subsection{Comparison to other XUV estimations}

\begin{figure}[t!]
\centering
\includegraphics[width=69mm]{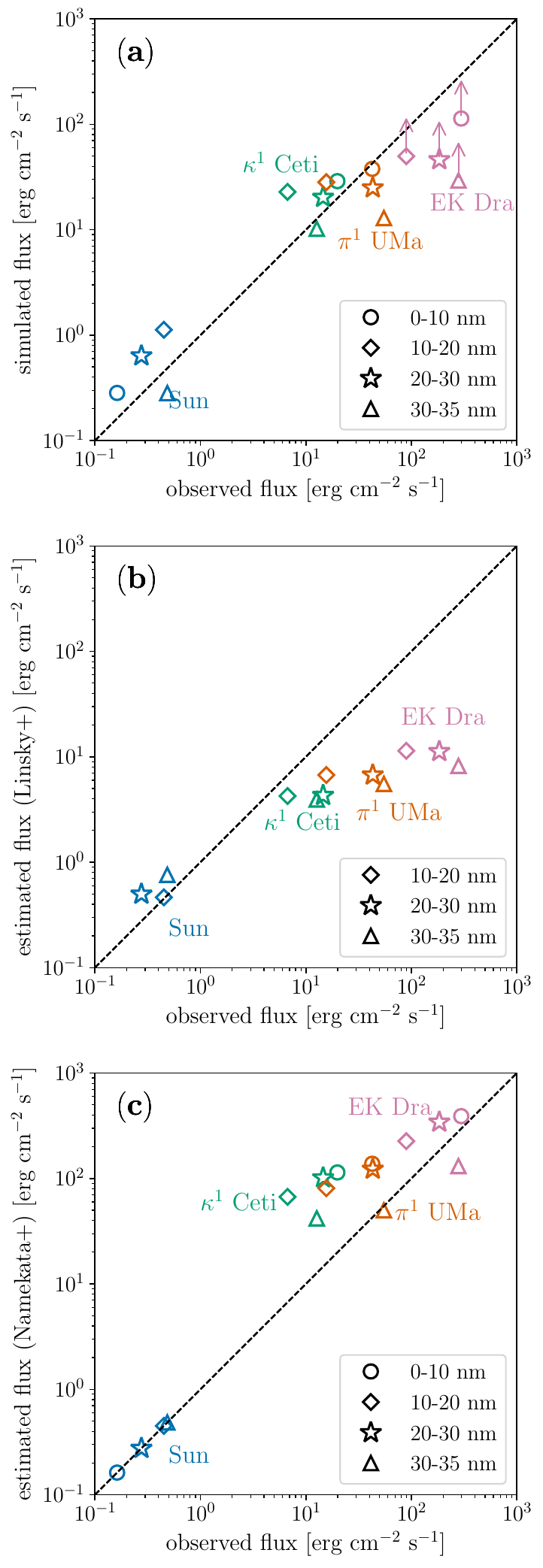}
\caption{Comparison of various XUV estimation methods and observational data. Following \citet{Linsky_2014_ApJ}, we divided the observable wavelength band into four bins and compared the flux in each bin, distinguished by symbols. Symbol colors correspond to different stars. (a) Comparison between our model and the observation. Arrows indicate the correction to the model spectrum due to insufficient resolution. (b) Comparison between the Ly-alpha-based estimatation \citep{Linsky_2014_ApJ} and observation. (c) Comparison between the estimation based on surface magnetic flux \citep{Namekata_2023_ApJ} and observation.}
\label{fig:XUV_bin_comparison}
\end{figure}

To evaluate the accuracy of our model in the XUV estimation, we compare its precision with other XUV estimation methods. Specifically, we compare our method against those using Lyman-alpha emissions \citep{Linsky_2014_ApJ} and surface magnetic flux \citep{Namekata_2023_ApJ} measurements.

To apply the Lyman-alpha based estimation, it is essential to accurately reconstruct the Lyman alpha line profile. This is because, like EUV, Lyman alpha is strongly absorbed by the interstellar medium, making it challenging to detect signals near the line center. The Lyman-alpha flux values for $\kappa^{1}$ Ceti and $\pi^1$ UMa are sourced from \citet{Linsky_2014_ApJ} and \cite{Ribas_2005_ApJ}, respectively. Since the Lyman alpha line profile of EK Dra was not reported before, we reconstructed it from the observed wing profile by using the open-source python code \textsf{lyapy}\footnote{\url{https://github.com/allisony/lyapy}} \citep{2022zndo...6949067Y}. We first obtained the FUV spectra including Lyman alpha line of EK Dra observed by the Hubble Space Telescope (HST) from the Mikulski Archive for Space Telescopes (MAST) archive. The data utilized was identified by the observation ID of oboq01010. EK Dra was observed in 2012 using the Space Telescope Imaging Spectrograph (STIS) instrument with the E140M filter through a 0.''2 × 0.''2 aperture. We integrated the total exposure time to obtain the spectrum of Ly$\alpha$. In reconstructing the spectral lines using \textsf{lyapy}, we set the spectral resolution to 50,000. As a result, the reconstructed intrinsic flux of the Lyman alpha was calculated to be 62.5 erg s$^{-1}$ cm$^{-2}$ at 1 au. 

It should be noted that Lyman-alpha-based estimations are subject to the inherent ambiguity associated with the reconstruction of the Lyman-alpha line profile. The discrepancy between observations and predictions may not necessarily stem from deficiencies within the model but could also reflect the inherent challenges associated with estimating Lyman-alpha flux.

For the estimation based on the surface magnetic flux, we utilize the XUV spectral data as presented in \citet{Namekata_2023_ApJ}. This data has a wavelength resolution of 0.1 nm and is available at: \url{https://github.com/KosukeNamekata/StellarXUV}.

Fig.~\ref{fig:XUV_bin_comparison} compares the accuracy of our model with the Lyman-alpha-based and magnetic-flux-based estimations against observational data. The Lyman-alpha-based estimation provides flux values in 10 nm bands. For a fair comparison, we calculate corresponding values for the other data sets. However, due to the absence of observational data above 36 nm, flux values are calculated in 30-35 nm instead of in 30-40 nm, and the Lyman-alpha-based estimation is halved in this band.

Lyman-alpha-based and magnetic-flux-based estimations show opposite trends in observational synthesis. As magnetic activity increases, Lyman-alpha-based estimations (Panel (b) of Fig.~\ref{fig:XUV_bin_comparison}) deviate more from observations, while magnetic-flux-based estimations (Panel (c) of Fig.~\ref{fig:XUV_bin_comparison}) become more aligned. Our model (Panel (a) of Fig.~\ref{fig:XUV_bin_comparison}), similar to Lyman-alpha-based estimation, tends to underestimate with increasing magnetic activity, but overall, its error margin is comparable or less than other methods. This comparison suggests that our method can reproduce the XUV emission spectra of solar-type stars with equal or greater accuracy than existing estimation techniques.

\subsection{Validation of DEM fitting via Chebyshev polynomials}

\begin{figure}[t!]
\centering
\includegraphics[width=75mm]{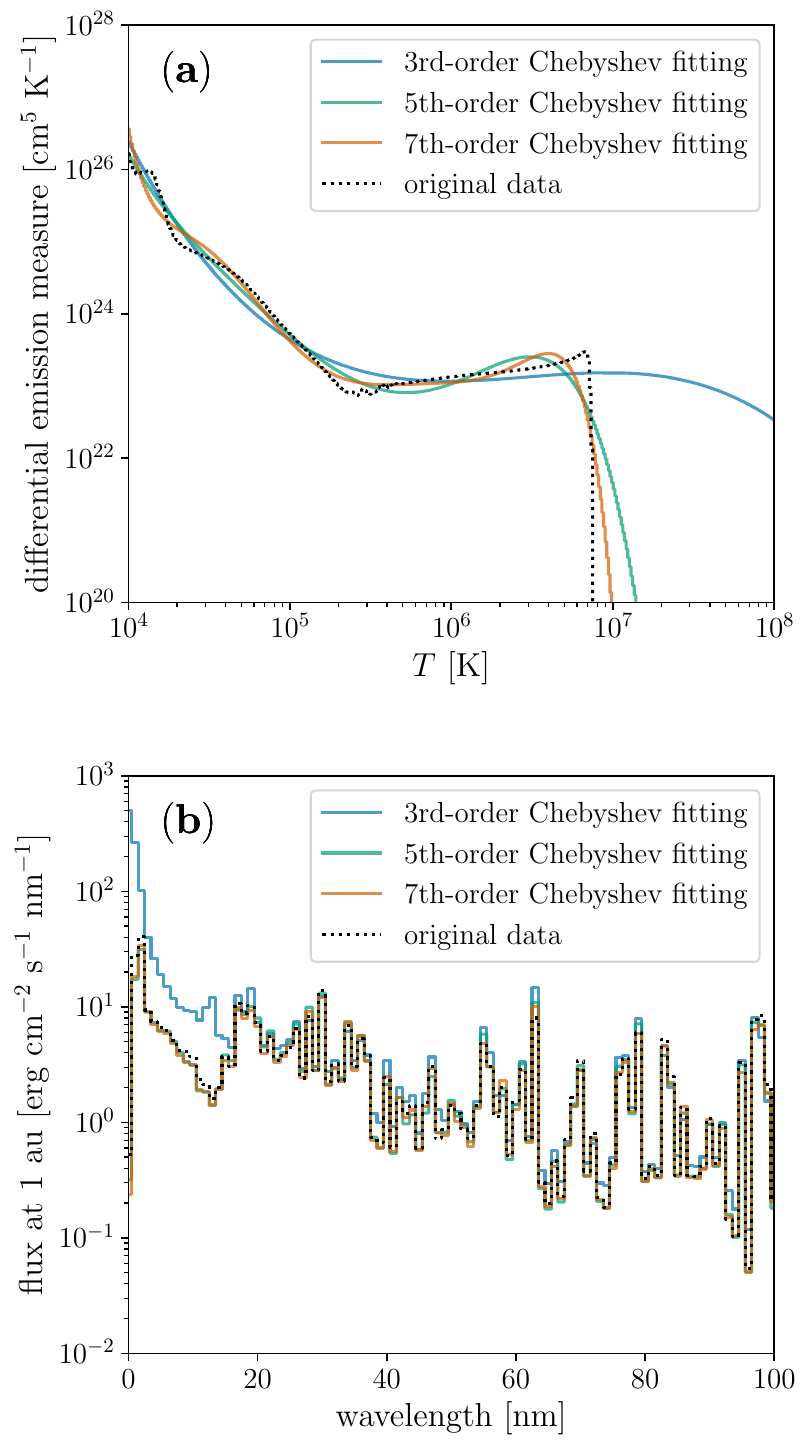}
\caption{Panel (a): Simulation-derived DEM for EK Dra (black dashed line) and its fitting using Chebyshev polynomials of degrees 3, 5, and 7 (solid lines). Panel (b): Computed XUV emission spectra using the DEM from Panel (a), with line styles matching those in Panel (a).}
\label{fig:Chebyshev_polynomial_comparion}
\end{figure}

In fitting the DEM obtained from ultraviolet and X-ray observations across a broad temperature range, we usually utilize fifth-order Chebyshev polynomials to fit the observed data. However, the empirical nature of this approach means the optimal polynomial degree for accurate DEM reconstruction is not self-evident. In this section, we examine the validity of using fifth-order Chebyshev polynomials for fitting, by fitting the model DEMs with polynomials of various degrees and analyzing the resultant spectra.

We employ the DEM of EK Dra as the dataset for fitting, focusing on data points within temperature ranges where DEM exceeds $10^{10} {\rm \ cm^{-5} \ K^{-1}}$. Fittings are performed in the log-log space with Chebyshev polynomials of orders up to $N$ as per the following equation \citep{Duvvuri_2021_ApJ}:
\begin{align}
    \log \xi (T) = \sum_{n=1}^N c_n T_n \left( \frac{\log T - 6}{2}  \right),
\end{align}
where $T_n$ are the $n$-th order Chebyshev polynomials of the first kind and $c_n$ are the corresponding coefficients.
$\xi (T)$ and $T$ are measured in units of ${\rm \ cm^{-5} \ K^{-1}}$ and ${\rm K}$, repsectively.
To examine convergence with respect to $N$, we test $N=3, \ 5, \ 7, \ 9$.

Panel (a) of Fig.~\ref{fig:Chebyshev_polynomial_comparion} displays the original DEM (shown as a black dashed line) alongside the fitting results. The 9th-order fitting is omitted for clarity. While the 3rd-order fitting significantly overestimates the DEM at higher temperatures ($T > 10^7 {\rm \ K}$), fittings of the 5th-order and above accurately capture the shape of the original data. Panel (b) of Fig.~\ref{fig:Chebyshev_polynomial_comparion} compares the bin-averaged XUV spectra derived from the fitted DEM to the original. The 3rd-order fitting overestimates the spectrum in the X-ray wavelength range, consistent with the overestimation of the DEM at high temperatures. In contrast, fittings up to 5th degree and higher exhibit high agreements. These results indicate that reconstructing DEM is feasible with 5th-degree or higher fittings, aligning with prior empirical findings.

To examine the quantitative error induced by the fitting with Chebyshev polynomials, we compare in Table 2 the X-ray and EUV luminosities ($L_{\rm X}$ and $L_{\rm EUV}$) derived from the original and fitted DEMs. Here, the X-ray wavelength range is defined as 0.517-12.4 nm, and the EUV wavelength range as 10-91.2 nm. Using fittings of fifth order or higher, the error in X-ray luminosity is at most $20\%$, and in EUV luminosity, at most $4\%$. We conclude that, in estimating XUV luminosity from DEM, fitting with Chebyshev polynomials is not likely to be the main source of error, as long as a 5th-order or higher fitting is employed and the fitted data are sufficiently accurate.

\renewcommand{\arraystretch}{1.5}
\begin{table}[t!]
\centering
  {\tabcolsep=1.2em
  \begin{tabular}{ccc}
    Case
    & $L_{\rm X}$ [erg s$^{-1}$]
    & $L_{\rm EUV}$ [erg s$^{-1}$]
    \\ \hline \hline
    3rd-order fitting 
    & $1.42 \times 10^{30}$ 
    & $7.79 \times 10^{29}$  \\
    5th-order fitting 
    & $2.58 \times 10^{29}$
    & $6.21 \times 10^{29}$ \\
    7th-order fitting 
    & $2.64 \times 10^{29}$
    & $5.90 \times 10^{29}$ \\
    9th-order fitting 
    & $3.09 \times 10^{29}$
    & $6.23 \times 10^{29}$ \\
    original data
    & $3.19 \times 10^{29}$ 
    & $6.12 \times 10^{29}$ \\
    \hline \hline
  \end{tabular}
  }
  \vspace{0.5em}
  \caption{
  Comparison of XUV luminosities calculated using the original DEM and the Chebyshev-polynomial-fitted DEMs.
  }
  \label{table:XUV_parameters_Chebyshev-polynomials}
\end{table}

\section{Summary and discussion \label{sec:discussion}}

In this study, we evaluate the capability and limitations of model-based estimates of stellar XUV emissions. Focusing on solar-type stars with both observed surface magnetic flux and metallicity (input parameters of the model) and XUV spectra (output parameter of the model), a comparison between model calculations and observations is conducted. The results demonstrate that our model accurately replicates observations and is as precise, if not more, than other estimation methods.

It should be noted that our model possibly fails to accurately describe the coronal heating. This is inferred from the fact that the average height of the transition region is 7.7 Mm during the solar maximum and 11.7 Mm during the solar minimum, both exceeding the conventional value \citep[$1.7-2.3 {\rm \ Mm}$,][]{Fontenla_1993_ApJ, Avrett_2008_ApJ}. One possible explanation for this significant discrepancy is that our model is one-dimensional, although we cannot rule out the possibility of overestimating energy injection from surface convection, implying that energy may be introduced through other mechanisms.

The most notable heating mechanism missing from our model is the impulsive heating associated with flares, which is considered particularly significant in young solar-type stars \citep{Audard_2000_ApJ, Ayres_2015_AJ}. We infer that flares are necessary to explain the X-ray spectrum below 1 nm, but in reality the effect of flares may be significant across the entire XUV wavelength range. To address this issue, it will be necessary to verify the findings using a coronal model that includes flare \citep{Cheung_2019_NatAs, Chen_2023_ApJ, Rempel_2023_ApJ}.

Additionally, heating from small loops associated with emerging magnetic fields \citep{Chitta_2020_AA, Chen_2021_AA, Wang_2022_SolPhys, Chitta_2023_ApJ, Pontin_2024_ApJ} and heating from coronal loop oscillations \citep{Anfinogentov_2015_AA, Nakariakov_2021_SSRev, Petrova_2023_ApJ, Lim_2023_ApJ, Lim_2024_arXiv} are not included. Further model validation, including these effects, is necessary. In any case, it cannot be concluded that the stellar corona is heated by turbulence solely because the observed stellar spectra can be approximately reproduced.

The turbulence model used in this study may require some modifications. We assume that the turbulence cascade efficiency, represented by the parameter $c_d$, remains constant regardless of the loop length. However, results from three-dimensional simulations show that the coronal heating rate strongly depends on the Alfv\'en crossing time, the time for an Alfv\'en wave to cross the coronal loop \citep{Rappazzo_2008_ApJ}. Therefore, including this effect will improve our turbulence model. This correction may be especially significant for stars with strong magnetic fields, where the Alfv\'en crossing time is expected to be much shorter than the Sun.

Despite the aforementioned limitations, our study suggests that numerical models, utilizing surface magnetic flux and elemental abundance as inputs, can serve as a powerful alternative method for estimating stellar XUV emissions. While focusing on solar-type stars in this study, our approach may extend to K- and M-type stars, abundant in exoplanets, and pre-main-sequence stars with protoplanetary disks, by adjusting boundary conditions appropriately. Applying our XUV estimation method to various low-mass stars could mitigate the uncertainty in XUV spectra, crucial for studying planetary atmospheres and protoplanetary disks.

\begin{acknowledgements}
Numerical computations were carried out on the Cray XC50 at the Center for Computational Astrophysics (CfCA), National Astronomical Observatory of Japan. In developing and optimizing the numerical code, we also utilized the computational resources of Wisteria/BDEC-01 Odyssey at the University of Tokyo, provided by the Joint Usage/Research Center for Interdisciplinary Large-scale Information Infrastructures in Japan (Project ID: jh230046). The authors thank Dr. Allison Youngblood for assistance with the usage of the python code \textsf{lyapy}. Some of the data presented in this paper were obtained from the Mikulski Archive for Space Telescopes (MAST) at the Space Telescope Science Institute. This work is supported by JSPS KAKENHI Grant Numbers JP21H04487, JP22KJ3091, JP22KK0043, JP22K14074, JP22K14077, and JP24K00688. This work made use of matplotlib, a Python library for publication quality graphics \citep{Hunter_2007_CSE}, and NumPy \citep{van_der_Walt_2011_CSE}.
\end{acknowledgements}

\bibliographystyle{aa}

\begin{appendix}

\section{$L_{\rm X}$--$L_{\rm EUV}$ scaling law}

In Paper I, we derived a scaling law between X-ray luminosity ($L_{\rm X}$) and EUV luminosity ($L_{\rm EUV}$). Given updates to our model since Paper I (see Section~\ref{sec:model}), and considering that $\kappa^1$ Ceti, $\pi^1$ UMa, and EK Dra exhibit different elemental abundances affecting radiative cooling functions and spectral calculation, the applicability of the Paper I scaling law to our current model is not guaranteed. For this reason, we test the $L_{\rm X}$-$L_{\rm EUV}$ scaling law from Paper I against our updated model.

To accurately assess the scaling law, we conduct additional simulations with hypothetical solar-type stars, possessing magnetic fluxes 3, 10, and 30 times that of the Sun (during its minimum activity phase), each with half-loop lengths of 80 Mm and 160 Mm. The solar coronal abundance was used for these simulations. With these six added simulations, we validate the scaling laws using a total of 27 computational results.

\begin{figure}[t!]
\centering
\includegraphics[width=75mm]{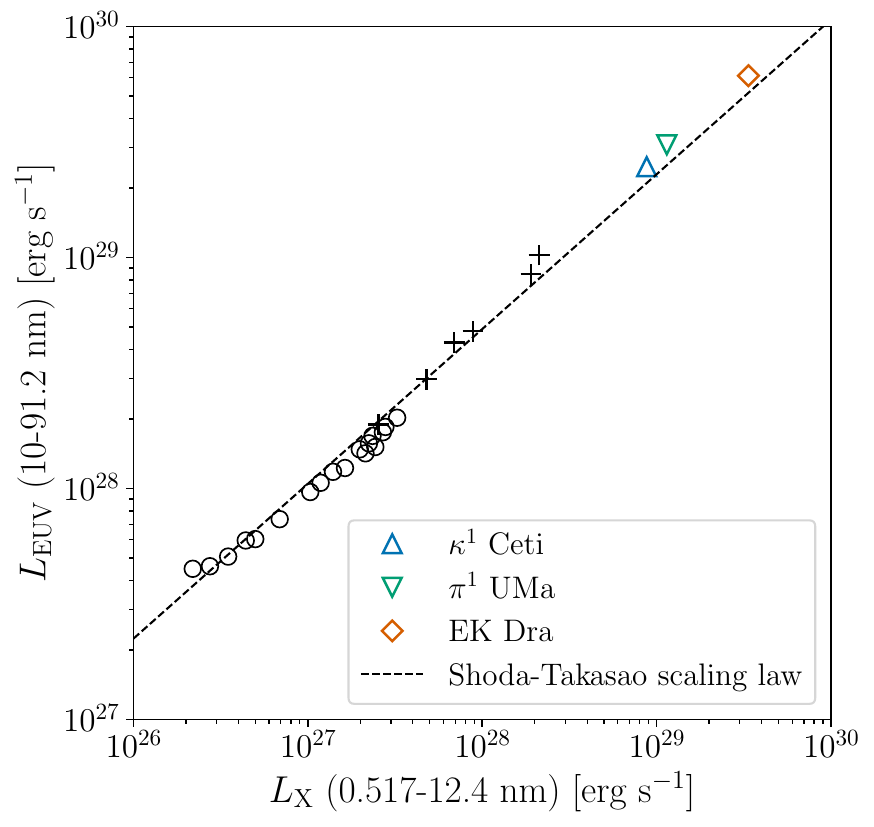}
\caption{Relationship between X-ray and EUV luminosity derived from this study. Colored symbols correspond to $\kappa^1$ Ceti, $\pi^1$ UMa, EK Dra. Circles correspond to the Sun, and pluses to hypothetical stars with 3-, 10-, and 30-times the solar magnetic flux. The dashed line shows the scaling law from Paper I. }
\label{fig:comparison_with_ST21_scaling_law_single}
\end{figure}

Fig.~\ref{fig:comparison_with_ST21_scaling_law_single} illustrates a comparison between the $L_{\rm X}$-$L_{\rm EUV}$ relationship obtained from our model and the scaling law presented in Paper I, hereafter referred to as Shoda--Takasao scaling law. Similar to the previous section, we define X-ray with wavelength ranging from 0.517 to 12.4 nm, and EUV wavelength from 10 to 91.2 nm. Although the X-ray wavelength range slightly differs from that used in Paper I, it has been verified not to impact subsequent discussions or conclusions. Shoda-Takasao scaling law is given as follows. 
\begin{align}
    \log L_{\rm EUV} = 9.93 + 0.67 \log L_{\rm X},
\end{align}
where $L_{\rm X}$ and $L_{\rm EUV}$ are measured in cgs unit. Fig.~\ref{fig:comparison_with_ST21_scaling_law_single} shows that this scaling law accurately agrees with the results of our model. Notably, the same scaling law applies to stars with slightly different atmospheric elemental abundances (and radiative cooling functions). This is possibly because as the radiative cooling function (XUV emissivity), $\Lambda (T)$, increases, the coronal density, $n_e$, decreases, leading to the XUV luminosity, $\propto n_e^2 \Lambda(T)$, becoming less dependent on $\Lambda (T)$ \citep{Washinoue_2019_ApJ, Washinoue_2023_ApJ}. Given the model's ability to precisely replicate the XUV emissions of highly active stars, it can be inferred that Shoda-Takasao scaling law is robust and applicable across a wide range of magnetic activities.

\section{Loop-length dependence of XUV spectrum for young solar-type stars \label{appendix:loop_length_dependence_piuma}}

This Appendix illustrates the loop-length dependence of XUV emission in the young solar-type star $\pi^1$ UMa. Given the extensive calculation time, we reduce the resolution slightly ($\Delta s_{\rm min} = 2 {\rm \ km}$) to facilitate a parameter survey.

\begin{figure}[t!]
\centering
\includegraphics[width=75mm]{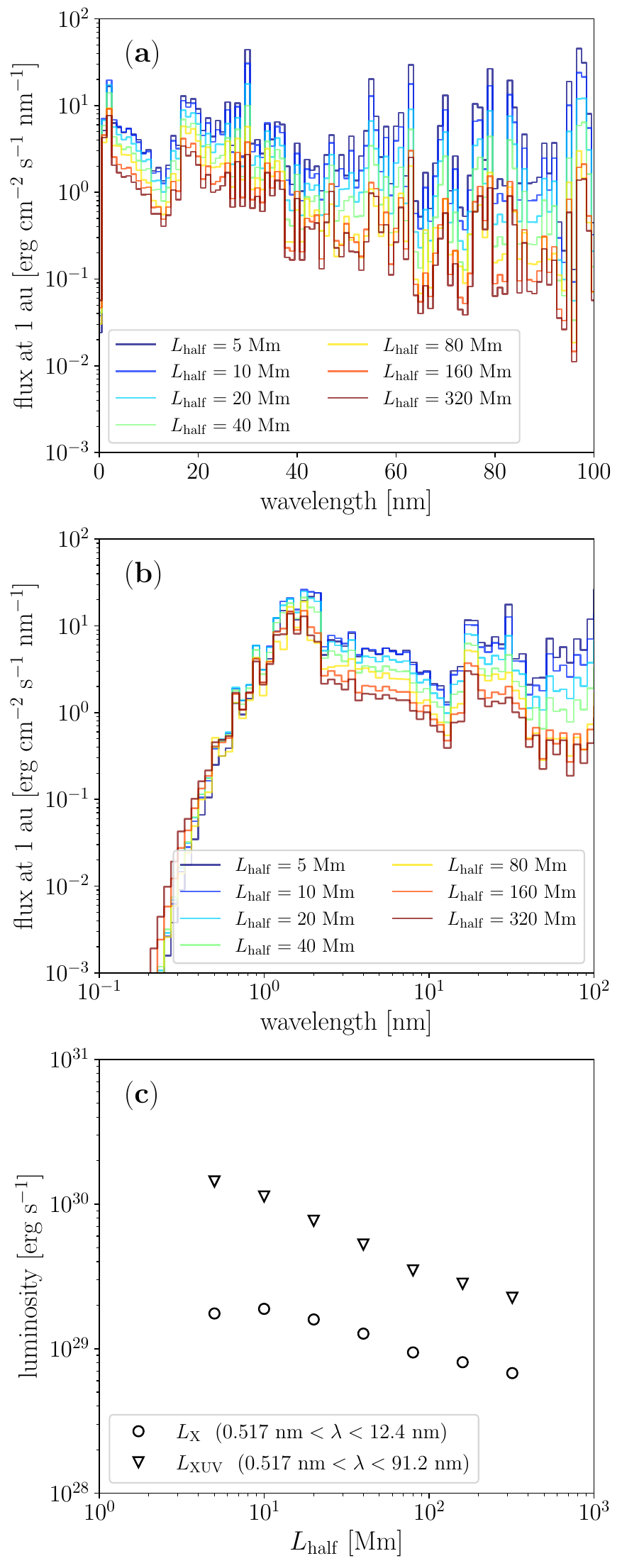}
\caption{Loop-length dependence of the XUV emission from $\pi^1$ UMa. Panels a and b: bin-averaged spectra, with different colors representing different loop lengths. The spectra were binned similarly to Fig.~\ref{fig:loop_length_dependence_solar_vertical}. Panel c: loop length dependence of X-ray and XUV luminosities.}
\label{fig:Loop_length_dependence_piuma_spectrum_dsmin2km_vertical}
\end{figure}

The top and middle panels of Fig.~\ref{fig:Loop_length_dependence_piuma_spectrum_dsmin2km_vertical} depict the relationship between the XUV spectrum and loop length. Notably, this dependence differs from that observed in the solar simulations (Fig.~\ref{fig:loop_length_dependence_solar_vertical}). Specifically, except for very short wavelengths (< 1 nm), shorter loops exhibit more intense XUV emission, contrary to the trend seen in the Sun. This suggests that in magnetically active stars, coronal heating is possibly more efficient in shorter loops, though the underlying reason remains unknown and requires further investigation. Furthermore, the dependence on loop length is more pronounced in the EUV range, also in contrast to the solar simulations.

The bottom panel of Fig.~\ref{fig:loop_length_dependence_solar_vertical} presents the X-ray and XUV luminosities versus loop length, with both luminosities decreasing as loop length increases. Therefore, in young solar-type stars like $\pi^1$ UMa, small-scale coronal loops may play a more significant role. While we currently lack methods to observe such small-scale magnetic structures on stars, acknowledging the potential importance of small-scale loops should be valuable for future studies.

\section{XUV estimation using the transition-region broadening technique \label{appendix:ltrac}}

One of the difficulties in coronal heating simulations is the underestimation of coronal density when the resolution in the transition region is insufficient \citep[transition-region problem,][see also Fig.~\ref{fig:resolution_dependence}]{Bradshaw_2013_ApJ, Shoda_2021_AA}. In studies where the quantitative estimation of radiation from the corona is crucial, this can be a critical issue.

One proposed method to address the transition-region problem is to expand its width. This method, which involves artificially altering the thickness of the transition region while maintaining energy balance through adjustments in radiative cooling and thermal conduction \citep{Lionello_2009_ApJ, Mikic_2013_ApJ, Johnston_2019_ApJ, Johnston_2021_AA_LTRAC, Iijima_2021_ApJ}, is quite effective. However, due to the broadening of the transition region along the line of sight, the DEM in the transition-region temperature range may be enhanced, potentially leading to a significant overestimation of the XUV spectrum.

Here, we quantitatively examine how broadening the transition region affects the DEM and the XUV spectrum calculated from it. The LTRAC method \citep{Iijima_2021_ApJ} is used for this purpose, with the free parameter set to the same value as in the original study ($p=4$). The validity of LTRAC is tested by comparing low-resolution calculations using the LTRAC method ($\Delta s_{\rm min} = 10 {\rm \ km}$) and high-resolution calculations without it ($\Delta s_{\rm min} = 1 {\rm \ km}$). Here, we compare the simulation results for $\kappa^1$ Ceti.

\begin{figure}[t!]
\centering
\includegraphics[width=75mm]{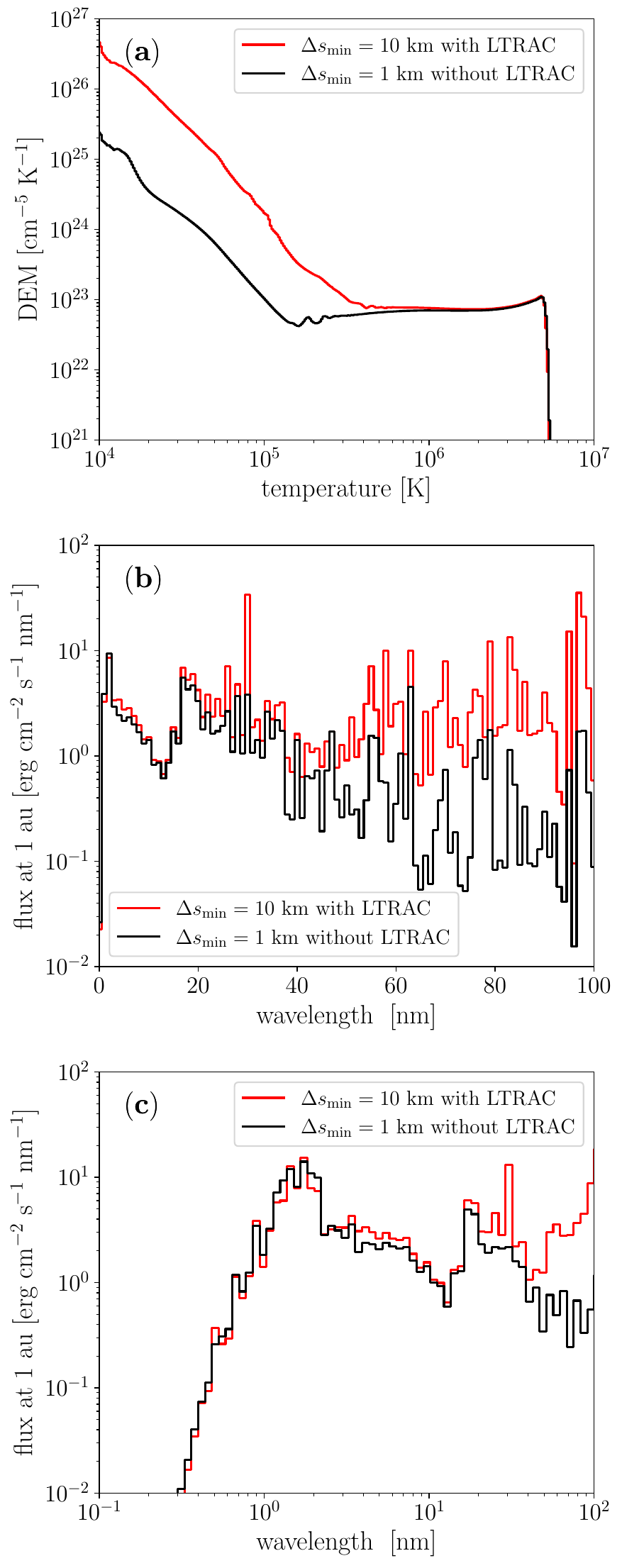}
\caption{Comparison between the low-resolution simulation using the LTRAC method and the high-resolution simulation without the LTRAC method. (a): DEMs (b) and (c): bin-averaged spectra. The binning of the spectra is performed in the same way as in Fig.~\ref{fig:loop_length_dependence_solar_vertical}.}
\label{fig:LTRAC_DEM_spectrum_comparison}
\end{figure}

Fig.~\ref{fig:LTRAC_DEM_spectrum_comparison} compares the low-resolution simulation using the LTRAC method (red lines) with the high-resolution simulation without the LTRAC method (black lines). As anticipated, the DEM significantly increases in the transition-region temperature range ($T \le 4 \times 10^5 {\rm \ K}$) with the LTRAC method. Meanwhile, the DEMs in the coronal temperature range ($T \ge 4 \times 10^5 {\rm \ K}$) remain nearly identical, demonstrating the effectiveness of the LTRAC method. Consequently, although the EUV emission (10-91.2 nm) is significantly overestimated using the LTRAC method, the X-ray emission (0.5-12.4 nm) can be accurately predicted even in the low-resolution simulation.

In summary, the LTRAC method is highly effective for predicting the X-ray spectrum but not applicable for predicting the EUV spectrum. While not directly confirmed, we anticipate the same result for other broadening techniques. To estimate the entire XUV spectrum using the LTRAC method, one possible approach could be to correct the DEM in the transition-region temperature range, which requires additional research.

\end{appendix}

\end{document}